\documentclass[
preprint,
 table , %
journal=jpcbfk, %
manuscript=article,%
,layout=twocolumn %
]{achemso}
\setkeys{acs}{maxauthors = 10,etalmode=truncate}

\usepackage{graphicx}%
\usepackage{graphbox}%
\usepackage{dcolumn}%
\usepackage{bm}%
\usepackage{comment}
\usepackage{siunitx}
\usepackage{pdfpages}
\usepackage{booktabs}
\usepackage{placeins}
\usepackage{subcaption}
\usepackage{layout}
\usepackage{xcolor} 
\usepackage{tikz}
\usetikzlibrary{backgrounds}
\usetikzlibrary{calc}
\usepackage{chemformula}
 \usepackage{braket}
\usepackage{textgreek}
\usepackage{academicons}
\usepackage{ifthen}
\usepackage{adjustbox}
\usepackage{academicons}
\definecolor{orcidlogocol}{HTML}{A6CE39}
\usepackage{multirow}
\usepackage[commandnameprefix=always,defaultcolor=orange,final]{changes}
\usepackage[switch,mathlines]{lineno}

\usepackage{soul}

\DeclareSIUnit{\calorie}{cal}
\DeclareSIUnit{\kcal}{\kilo\calorie}
 \DeclareSIUnit\Debye{D}
 \DeclareSIUnit\angstrom{\text {Å}}
 \DeclareSIUnit\bar{bar}
\usepackage{pgffor}

\newcounter{pdfpages}

\usepackage{amssymb,amsmath}

\DeclareMathOperator{\Debye}{D}

\usepackage{chemfig}

\usepackage[linesnumbered, ruled]{algorithm2e}
\SetKwRepeat{Do}{do}{while}%
\SetKwProg{Fn}{Function}{}{}%
\SetKwInput{Col}{Collaborators}{}%
\SetKwInput{Resp}{Responsibility}{}%

\author{René Hafner}
\affiliation{Physics Department and Research Center OPTIMAS, University Kaiserslautern-Landau, Erwin-Schrödinger-Straße, 67663 Kaiserslautern, Germany}%
\alsoaffiliation{Fraunhofer ITWM, Fraunhofer-Platz 1, 67663 Kaiserslautern, Germany}
\author{Nils Wolfgramm}
\affiliation{Fraunhofer ITWM, Fraunhofer-Platz 1, 67663 Kaiserslautern, Germany}
\author{Peter Klein}%
\email{peter.klein@itwm.fraunhofer.de}
\affiliation{Fraunhofer ITWM, Fraunhofer-Platz 1, 67663 Kaiserslautern, Germany}
\author{Herbert M. Urbassek}%
\affiliation{Physics Department and Research Center OPTIMAS, University Kaiserslautern-Landau, Erwin-Schrödinger-Straße, 67663 Kaiserslautern, Germany}
\email{urbassek@rhrk.uni-kl.de}

\title{Adsorption of Diclofenac and PFBS on Hair Keratin Dimer}
\usepackage{xr} %

\usepackage[hypertexnames=false,colorlinks, citecolor = blue, linkcolor=blue, urlcolor=blue]{hyperref}
\usepackage[noabbrev,nameinlink,capitalise]{cleveref}
\Crefname{equation}{eq}{eqs}
    \makeatletter
    \newcommand*{\addFileDependency}[1]{
      \typeout{(#1)}
      \@addtofilelist{#1}
      \IfFileExists{#1}{}{\typeout{No file #1.}}
    }
    \makeatother

\SectionNumbersOn
\begin{document}

\date{\today}%
\begin{abstract}
Environmental pollution by man-made toxic and persistent organic compounds, found throughout the world in surface and groundwater, has various negative effects on aquatic life systems and even humans. Therefore, it is important to develop and improve water treatment technologies capable of removing such substances from wastewater or purifying drinking water. The two substances investigated are the widely used painkiller diclofenac and a member of the class of "forever chemicals", perfluorobutane sulfonate. Both are known to have serious negative effects on living organisms, especially under long-term exposure, and are detectable in human hair, suggesting adsorption to a part of the hair fiber complex. 
In this study, a human hair keratin dimer is investigated for its ability to absorb diclofenac and perfluorobutane sulfonate. Initial predictions for binding sites are obtained via molecular docking and subjected to molecular dynamics simulations for more than $1$~\si{\micro\second}. The binding affinities obtained by the linear interaction energy method are high enough to motivate further research on human hair keratins as a sustainable, low-cost, and easily allocatable filtration material.
\end{abstract}

\maketitle

\section{Introduction}

The levels of pollution by humans through the release of organic compounds into the environment are still increasing, leading to long-term exposure to organisms, including humans. Toxic and long-lived organic compounds, such as perfluoroalkyl and polyfluoroalkyl substances (PFAS), and many drugs, are found in particular in aqueous environments around the world according to recent studies \cite{Kurwadkar.2022,Wilkinson.2022}. Both groups of substances can cause multiple serious biological defects even in offspring of exposed individuals\cite{Li.2022,McCarthy.2021, Sinclair.2020, Dickman.2022}. Therefore, it is of paramount importance to advise on technologies for the treatment of (waste) water streams contaminated by such substances in typically low concentrations; this is still an active field of research as, for example, reviewed in refs~\citenum{Yadav.2022, Akintola.2023,Shearer.2022}.

In this work, two typical examples from the group of PFAS and drug substances are investigated: perfluorobutane sulfonate (PFBS) and diclofenac (DIC). PFAS substances are durable under standard environmental conditions and, therefore, belong to the so-called ``forever chemicals". Short--chain PFBS is chosen because of its increasing use as an alternative to perfluorooctane sulfonate (PFOS). Its toxicity is shown in particular in refs \citenum{Karnjanapiboonwong.2018,Chowdhury.2021}. DIC is a non-steroidal anti-inflammatory human and veterinary drug. There is evidence on the bioaccumulation and bioconcentration of DIC in marine water, sediment and organisms \cite{Bonnefille.2018}, and in avian food above recommended thresholds \cite{Peters.2022}.

The treatment of contaminated water requires sustainable, low-cost, and globally available filtration materials. Human hair keratins were chosen in this study because of their high chemical and mechanical stability, and their charged nature, which should favor adsorption for both molecules since they are in their anionic form at standard pH value, see below.

A keratin based filter may consist of a membrane of intertwined hair fibers, while even a single hair exhibits a highly ordered structural hierarchy, including a cysteine-rich protein matrix and pigments such as melanin.\cite{Yang.2014} As a first step, a single keratin dimer is modeled as it may be present and accessible in such a matrix. Biomonitoring studies have shown that human hair can act as an indicator of drug usage and pollutant exposure for both diclofenac\cite{Gaillard.1997,Rothe.1997} and PFBS\cite{Alves.2015,Piva.2021}, among many other pharmaceuticals and PFAS. This suggests that both molecules interact with an as yet unknown part of the hair fiber. Colored hair contains small pigments that are variants of the polymer melanin. Many drugs are known to interact with melanin\cite{Ings.1984}, however, it was recently shown that diclofenac binds only weakly to these polyanionic polymers.\cite{Reinen.2018} Furthermore, \citeauthor{Rodriguez.2020b}\cite{Rodriguez.2020b} produced activated carbon from tennary hair waste, which exhibited adsorption capability for diclofenac. These indications motivate the study of the molecular level interaction of keratins with diclofenac and PFBS as model compounds.

 Therefore, the potential of the treatment of water streams containing PFBS and DIC by adsorption on hair proteins is explored using a molecular docking and a subsequent molecular dynamics simulation approach. %
\section{Theory and Computational Details}
\label{sec:methods}

In this section, the molecules, the protein heterodimer, and the docking and molecular dynamics (MD) protocols are presented. Furthermore, the details of the interaction fingerprint and energy analysis are introduced.

\subsection{Molecular models of DIC and PFBS}

\begin{figure}[t]
 \begin{subfigure}[t]{1.5in}
\caption{DIC}
	\includegraphics[align=c,width=\linewidth]{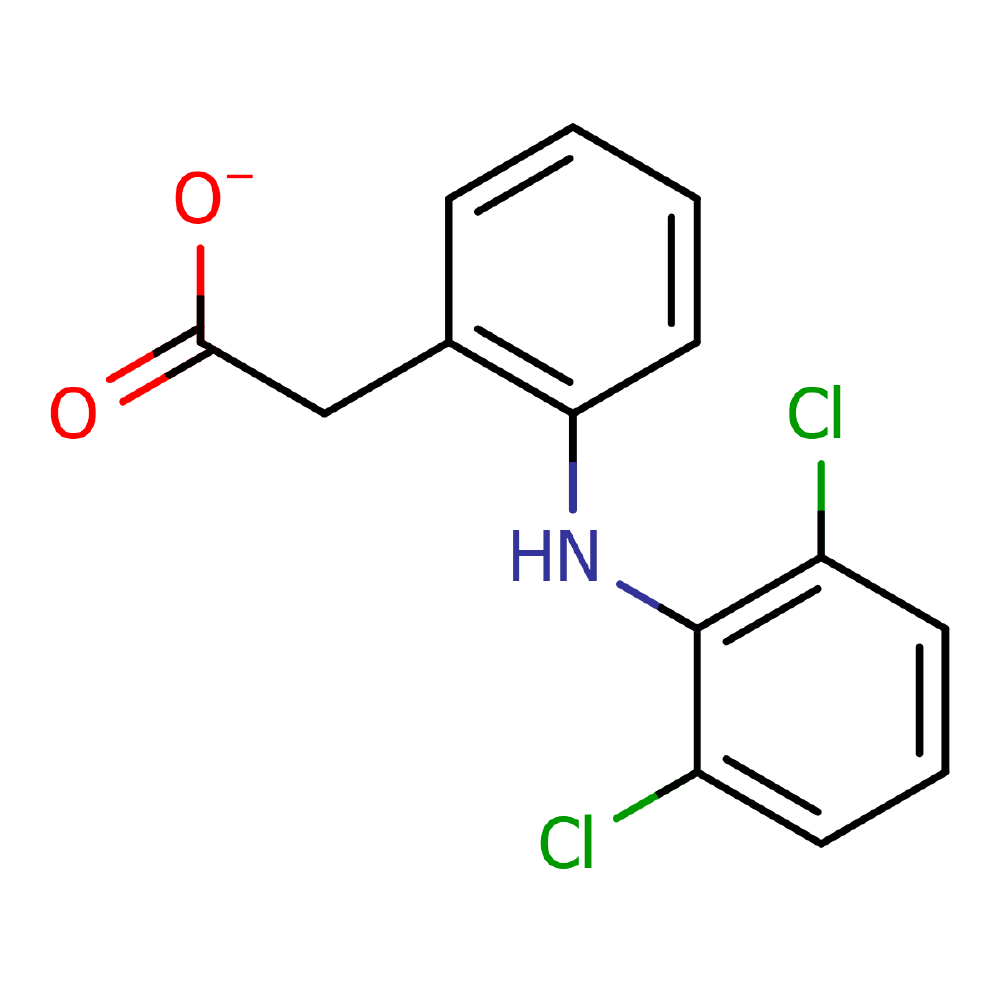}
			\label{fig:dic_structure}
	\end{subfigure}
	\begin{subfigure}[t]{1.5in}
 \caption{PFBS}
\vspace*{1cm}
 \includegraphics[align=c,width=\linewidth]{{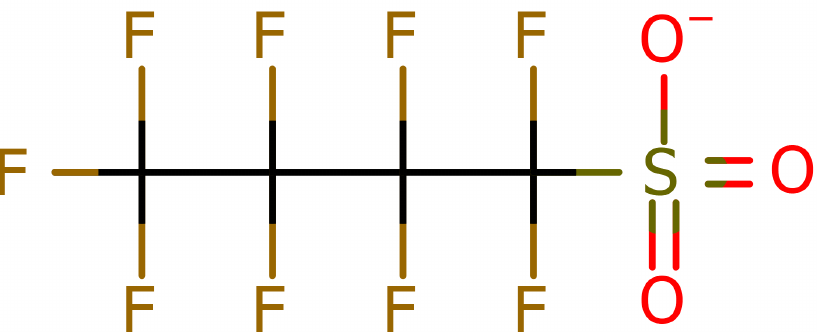}}

			\label{fig:pfbs_structure}
	\end{subfigure}
	\caption{Two--dimensional structures of (a) diclofenac and (b) perfluorobutane sulfonate.}
 \label{fig:molecule_structures}
\end{figure}

The chemical structures of DIC and PFBS are shown in \cref{fig:molecule_structures}. DIC (molecular weight $m_w= 296.1$~\si{\gram\per\mol}) contains two phenyl groups connected by a nitrogen atom. A carboxyl group and two chlorine atoms are attached to the phenyl groups, respectively. PFBS ($m_w= 300.09$~\si{\gram\per\mol}) consists of a short fully fluorinated carbon chain and a sulfonic acid end group. The Simplified Molecular Input Line Entry Specification (SMILES) strings of the ligands are listed in\setcitestyle{numbers}refs \bibnote{Diclofenac SMILES string: {\footnotesize c1c(Cl)c(c(Cl)cc1)Nc1c(cccc1)CC(=O)[O-]}} and \bibnote{PFBS SMILES string: {\tiny FC(F)(F)C(F)(F)C(F)(F)C(F)(F)S(=O)(=O)[O-]}}.\setcitestyle{super} 

The charge state of the ligands was investigated using the chemicalize webapp\bibnote{Chemicalize was used for the charge state prediction, 05/2019, https://chemicalize.com/, developed by ChemAxon} 
reporting a pKa value of $4.2$ and $-0.3$ for DIC and PFBS, respectively. 
Therefore, at pH $7$, both ligands are modeled with a negative charge on their carboxyl and sulfonate groups, respectively. 

For both ligands, the Charmm General Force Field (CGENFF) v3.0.1\cite{Vanommeslaeghe.2010,Vanommeslaeghe.2012,Vanommeslaeghe.2012b} is used, with missing parameters generated using the ParamChem server\cite{paramchemwebserver} and made freely available, see ref \citenum{K35-K85-molmod}.

Since multiple ligands and protein pockets are studied, different ligands and sites are distinguished by a prefix DOX or POX, where the first letter represents the ligand (DIC or PFBS) and X is a number.

\subsection{Human hair keratin heterodimer modeling}
\label{sec:methods-dimer}

The target adsorbent to be investigated  are human hair fibers, that on the most basic level consists of keratin protein heterodimers.\cite{Wang.2016c} In general, keratins are categorized in type I (acidic) and type II (neutral to basic) groups.\cite{Langbein.1999,Langbein.2001,Langbein.2007} 
 A pair of type I and type II proteins can form a heterodimer with a coiled coil structure, i.e., two $\alpha$ helices wrapped around each other.  
 However, only certain pairs of type I and type II proteins actually form a heterodimer. In this work, the nomenclature introduced in ref \citenum{Schweizer.2006} is followed: The human hair keratins modeled in the following are named K31-K40 and K81-K86, for types I and II, respectively.

The literature and protein database were scanned for models of human hair keratin heterodimers; however, no experimental study of any complete atomistic structure of any hair keratin heterodimer was found.

 Homology modeling is one successful approach to obtain models for intermediate filament proteins due to their structural similarity. On the other hand, ref.~\citenum{Chou.2012} used two coiled coil structure prediction algorithms to build heterodimer and tetramer keratin structure models. They report a complete atomistic heterodimer structure formed by the acidic K35 and the basic K85 protein. The amino acid sequences correspond to the UniProt sequences Q92764 and P78386, respectively\cite{k35structureRECENT,k85structureRECENT}.
 This atomistic model of a heterodimer is used as a starting point in this work.

Some structural issues were detected in the PDB file provided in ref \citenum{Chou.2012}. The program VMD\cite{Humphrey.1996} and its plugins Chirality and Cispeptide\cite{Schreiner.2011} were used to remove detected cis peptide bonds and chirality %
issues within the protein heterodimer structure.
The charge states of the protein residues were investigated with DelphiPka\cite{Wang.2015,Wang.2016,Pahari.2018} and  no deviations
to standard charges were found for the basic (ARG, LYS) and acidic residues (ASP, GLU). 

Histidine, which is present 6 times in the heterodimer, is a special amino acid in that its $\mathrm{pKa} = 6.2$, which means that at pH $7$, following the Henderson–Hasselbalch equation\cite{Gold2019},
approx. $10\%$ of an ensemble should be present in its protonated form. 
However, single residue titration revealed a protonated fraction of less than $21\%$ for each of the six histidines present in the heterodimer, but for residue HIS235 in segment 1B of monomer K35 with a value of $40\%$. 

As the major focus in this work is on the terminal regions without any histidine present, this will not influence the findings of this study.

The CHARMM36 protein force field with CMAP correction was used for the protein.\cite{MacKerell.1998,Huang.2013,Huang.2017} The CHARMM force field was chosen as it was used in the original publication of the structure\cite{Chou.2012} and successfully used in the measurement of binding affinities between epithelial keratin and other ligands.\cite{Marzinek.2014PhDThesis,Marzinek.2013,Zhao.2013,Zhao.2014}

\subsection{Equilibration Protocol}
\label{subsec:equiprotocol}

Simulation and post-processing runs are performed using NAMD2.14\cite{Phillips.2005,Phillips.2020} if not stated otherwise.

The simulation inputs in this study were created using the QwikMD\cite{Ribeiro.2016} interface in VMD\cite{Humphrey.1996}.
Systems are solvated using the explicit TIPS3P water model\cite{MacKerell.1998,Martyna.1994}, neutralized and additional ions are added to achieve a salt concentration of $0.15$~\si{\mol\per\litre}. In all equilibration simulations \ch{NaCl} salt was used.
The ligands are generally used in experiments in their salt form, i.e., diclofenac sodium and PFBS potassium. Therefore, for the production molecular dynamics simulation, systems containing the ligand DIC and PFBS, the sodium or the potassium ion was selected as the cation, respectively.
Short-range, non-bonded interactions are calculated with a distance cutoff of $12$~\si{\angstrom}. The van--der--Waals (vdW) interactions are smoothed out starting at $10$~\si{\angstrom}. The long range electrostatic interaction were treated by the particle--mesh Ewald (PME)\cite{Darden.1993} method.

An integration timestep of $2$~\si{\femto\second} was used with short--range interactions evaluated at every step and long--range interaction at every second step. Pressure is maintained at $1$~atm using the Nosé-Hoover Langevin piston method\cite{Martyna.1994,Feller.1995} using standard values with a postion period of $200$~\si{\femto\second} and a piston decay time of $100$~\si{\femto\second}. The temperature is controlled by Langevin dynamics\cite{Bruenger.1984} at $300$~\si{\kelvin} with a damping coefficient of $1$~\si{\per\pico\second}.

The following equilibration protocol was employed:
First, a minimization is conducted in $2000$ steps, with the atoms of the protein backbone restrained. The second phase is the annealing phase, with the temperature initialized at $60$~\si{\kelvin} and elevated to $300$~\si{\kelvin} at a rate of $1/1.2$~\si{\kelvin\per\pico\second}. The system is further relaxed in the NpT ensemble for $1$~\si{\nano\second} at $300$~\si{\kelvin} and $1$~\si{\bar}, still keeping the protein backbone restricted.
Next, full dynamics was enabled for at least $50$~\si{\nano\second} in the NpT ensemble, by removing the restriction on the protein backbone atoms.

In order to find possible interaction sites between the two ligands and the protein, a two-step approach using molecular docking followed by molecular dynamics (MD) was employed.

\subsection{Docking}
Docking algorithms are generally used to find one or multiple binding sites on proteins that are specific to a molecule to initiate, inhibit, or accelerate a reaction. Docking analysis requires high-resolution crystallographic data and/or well-equilibrated molecular structures as input, which are obtained in this work using the equilibration protocol described in \cref{subsec:equiprotocol}.

 AutoDock VINA\cite{Trott.2010} (version 1.1.2) was used to obtain and investigate binding site predictions. While docking programs in general are great tools for determining the geometric orientation or pose of a ligand in a binding site, the correlation coefficient of the docking score returned by AutoDock VINA and the experimental binding affinity is rather limited ($\approx 0.493
$)\cite{Pham.2022,Wang.2016b}, but comparable to other docking programs.\cite{Wang.2016b}

Due to the large extent of the K35/K85 keratin heterodimer and the presence of charged residues in its sequences, it is expected that more than a single site per heterodimer will be found.
The VINA program places a ligand of interest on a grid and uses a quasi-Newtonian optimization method to quickly find a local minimum in a predefined volume using a scoring function and its gradient as inputs. The scoring function consists of two distance-dependent Gaussian functions, one repulsion term, one term describing hydrophobic repulsion, and another term for hydrogen bonding using empirical weights. See refs \citenum{Trott.2010,Quiroga.2016} for their exact definitions. For a single run, the best pose and up to eight local minima are reported for the defined search volume.
Since the search space of VINA for a single run is limited to $30\times30\times30$~\si{\cubic\angstrom} by default, the program is started several times so that each residue is contained at least once in a search volume. The number of runs required, then depends on the size of the protein or segment under investigation. Selected results of the top (high affinity) ligand poses are then combined into a single pdb file for further usage.

\subsection{Molecular dynamics adsorption modeling}

\begin{figure}[htb]
	\begin{subfigure}[t]{\linewidth}
  \includegraphics[width=\linewidth]{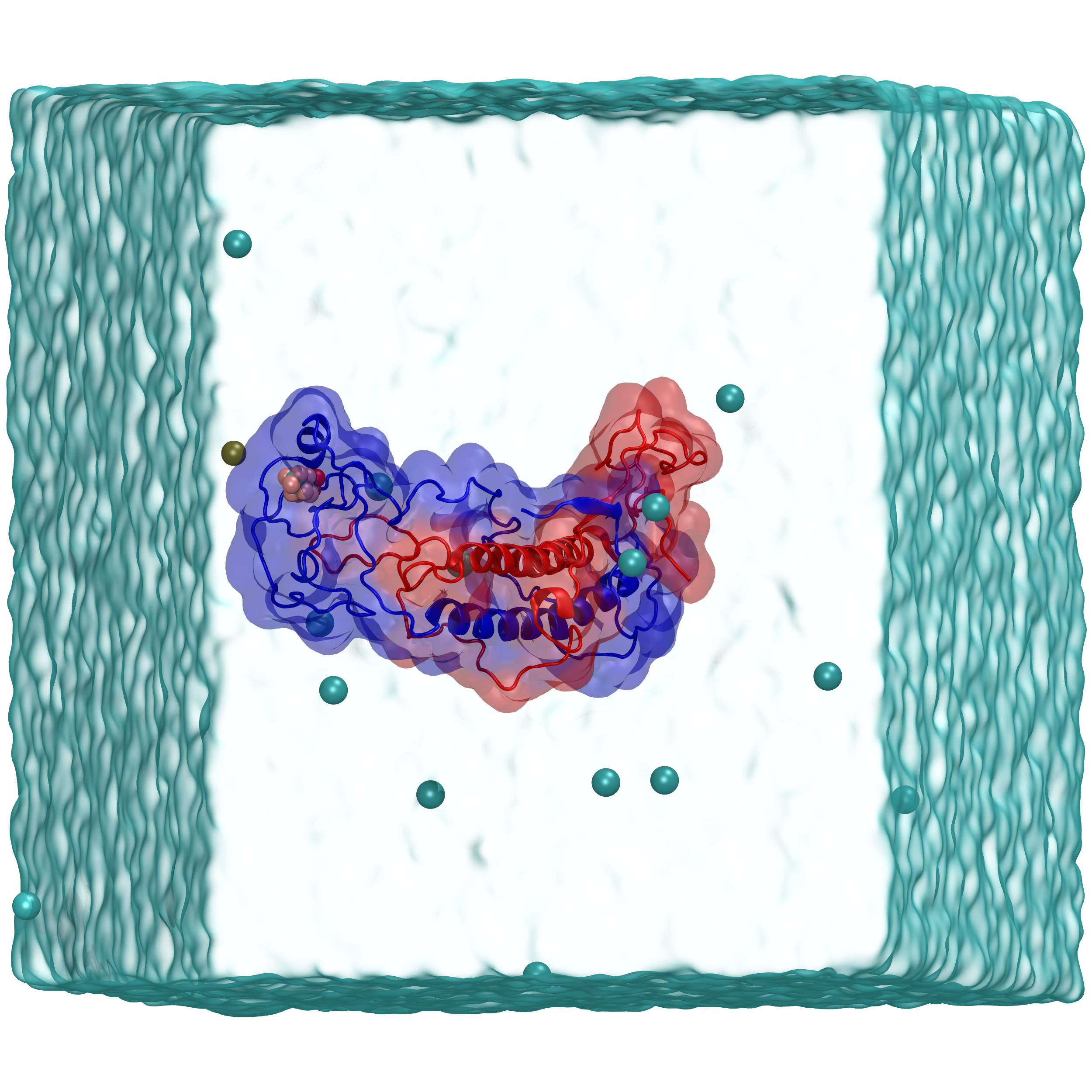}
\end{subfigure}
	\caption{Illustration of the setup used in molecular dynamics production runs for single ligand adsorption.}
	\label{fig:system_illustration}
\end{figure}

Structures obtained by docking are sorted according to their VINA score and the most promising binding sites of ligands, including their poses, are used to generate input structures for molecular dynamics runs. 

To this end, a truncated heterodimer is used to save computational expenses; see \cref{sec:results} for further details. 
The docking structures are extended by including a water environment, including salts and counter ions to make the total system electrically neutral. 
Then, these newly constructed molecular systems of heterodimer fragments and ligands dissolved in water containing salt ions are equilibrated using the simulation setup described in \cref{subsec:equiprotocol}. 

In molecular dynamics production runs, the equilibrated systems were then propagated at fixed temperature and pressure.
The generated trajectories were then analyzed as detailed below in order to characterize the binding affinity of the ligands. This approach has a higher predictive power than docking and thus leads to a quantitatively better characterization of binding characteristics. 
In this work, molecular interaction fingerprints and Gibbs free energy estimations are used as described below. 

An example setup for molecular dynamics production runs is shown in \cref{fig:system_illustration}. A single PFBS molecule adsorbs on the heterodimer with counter ions present to neutralize the system. Further details are given in the results, \cref{sec:results}.

\subsection{Molecular interaction fingerprints}

The ProLIF Python library\cite{Bouysset.2021}, based on the rdkit library\cite{GregLandrum.2023}, was used to find the molecular interaction fingerprint along the MD trajectories of the ligands with the heterodimer, i.e., to identify key interactions with the protein and the types of residues that bind the ligand to the protein surface. 
Analysis is carried out based on a search along the trajectory in a distance-based manner using SMILES arbitrary target specification (SMARTS) patterns\cite{SMARTpatternsWebsite}. The types of interaction available are vdW contacts, hydrophobic (contacts between nonpolar moieties), anionic (anionic ligand with a cationic protein residue), $\pi$--stacking or $\pi$--$\pi$, $\pi$--cation, and hydrogen bonding. Furthermore, halogen bonds are treated similarly to hydrogen bonds. The standard distance and angle cutoffs are used in this work.
This allows for an in-depth analysis of residue type statistics and the examination of the continuity of specific ligand--residue interactions by creating interaction timelines.

\subsection{Gibbs free energy methods}

Two common approaches to compute the Gibbs free energy of binding along MD trajectories are end-point free energy methods, like molecular mechanics Poisson-Boltzmann surface area (MM/PBSA)\cite{Genheden.2015,Wang.2017}, and the linear interaction energy (LIE) model.\cite{Hansson.1998}

In this study, the updated standard LIE method is used. The Gibbs free energy of adsorption of a ligand in a bound complex is estimated by weighting the vdW and electrostatic (EL) contributions to the interaction energy difference compared to a ligand in bulk water in the following way:
\begin{equation}
\label{eq:model_lie_s}
\begin{split}
\Delta G =&~~~\alpha (\langle E_{\mathrm{vdW}}^{\mathrm{bound}} \rangle - \langle E_{\mathrm{vdW}}^{\mathrm{free}} \rangle) \\
& + \beta (\langle E_{\mathrm{EL}}^{\mathrm{bound}} \rangle -\langle E_{\mathrm{EL}}^{\mathrm{free}} \rangle) \\
    =&~ \alpha\cdot \Delta E_{\mathrm{vdW}}+\beta\cdot \Delta E_{\mathrm{EL}},
\end{split}
\end{equation}
where $\alpha=0.18$ and $\beta=0.5$\cite{Hansson.1998}.
The value of $\beta$ is the result of the linear response theory\cite{Aqvist.1994}, while the parameter $\alpha$ is obtained from fitting experimental data to computational results. %
In the absence of experimental values for keratins no optimization of the $\alpha$ parameter is conducted and the standard value is used. Thus, \cref{eq:model_lie_s} describes a simple estimate for the binding affinity for protein-ligand systems. However, the LIE model expects a pair-wise additive calculation for the energies. Therefore, instead of utilizing PME, in the energy extraction step the electrostatic energy is calculated using a switching and cutoff scheme the same way as for the vdW energy with an increased switching distance of $15$~\si{\angstrom} and a cutoff of $16$~\si{\angstrom}.

\section{Results and Discussion}\label{sec:results}

\begin{figure*}[tb]
 \begin{subfigure}[t]{\linewidth}
     \phantomsubcaption
     \label{fig:keratinsegments}
 \end{subfigure}
  \begin{subfigure}[t]{\linewidth}
     \phantomsubcaption
     \label{fig:keratinchargedresidues}
 \end{subfigure}
 \hspace*{-0.25cm}
\begin{subfigure}[t]{\linewidth}
 \begin{tikzpicture}[transform shape,rotate=0]%
\usetikzlibrary{positioning}
	\tikzstyle{every path}=[line width=2pt]
	\node[anchor=south west,inner sep=0] (image) at (0,0) {
 \includegraphics[width=\linewidth,clip,trim={0px 0 200px 0}]{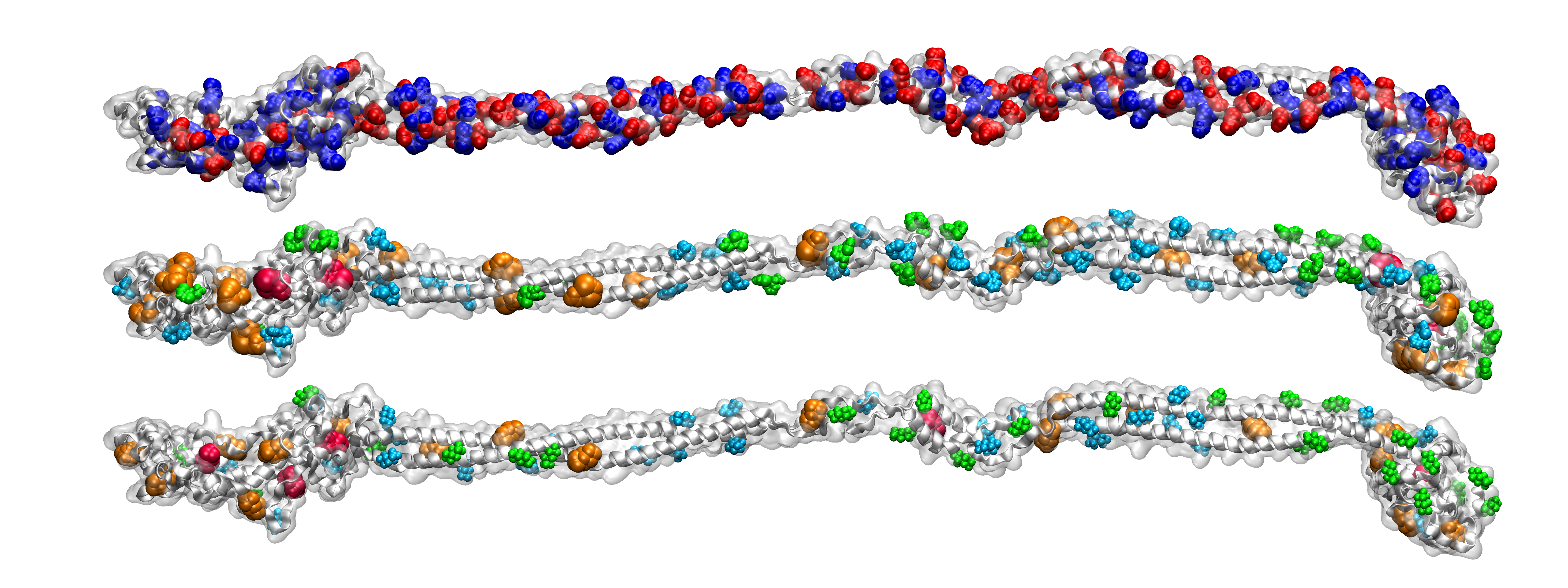}};
\begin{scope}[x={(image.south east)},y={(image.north west)}]
\draw node[shape=rectangle,minimum size=0.3cm,anchor=west,rotate=0,fill=none] (c) at (0.015,0.85) {(b)};
\draw node[shape=rectangle,minimum size=0.3cm,anchor=west,rotate=0,fill=none] (c) at (0.015,0.625) {(c) DIC};
\draw node[shape=rectangle,minimum size=0.3cm,anchor=west,rotate=0,fill=none] (c) at (0.015,0.35) {(d) PFBS};
\draw node[shape=rectangle,minimum size=0.3cm,anchor=west,rotate=0,fill=none] (c) at (0.015,0.95) {(a)};
\draw node[shape=rectangle,minimum size=0.3cm,anchor=west,rotate=0,fill=none] (c) at (0.1,0.95) {HEAD  1A};
\draw node[shape=rectangle,minimum size=0.3cm,anchor=west,rotate=0,fill=none] (c) at (0.25,0.95) {L1};
\draw node[shape=rectangle,minimum size=0.3cm,anchor=west,rotate=0,fill=none] (c) at (0.35,0.95) {1B};
\draw node[shape=rectangle,minimum size=0.3cm,anchor=west,rotate=0,fill=none] (c) at (0.625,0.95) {L12};
\draw node[shape=rectangle,minimum size=0.3cm,anchor=west,rotate=0,fill=none] (c) at (0.75,0.95) {2};
\draw node[shape=rectangle,minimum size=0.3cm,anchor=west,rotate=0,fill=none] (c) at (0.9,0.95) {TAIL};
 	\end{scope}
\end{tikzpicture}

\phantomsubcaption
\label{figsub:vina_results_distribution_image_a}
\end{subfigure}

\vspace*{-1.5cm}

 \begin{subfigure}[t]{3in}
  \includegraphics[width=\linewidth,page=1]{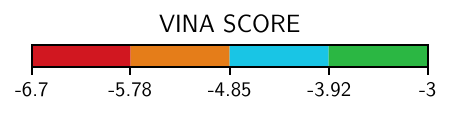}
\phantomsubcaption
\label{figsub:vina_results_distribution_image_b}
\end{subfigure}
\vspace*{-0.5cm}
\begin{subfigure}[]{\linewidth}
 \begin{tikzpicture}[transform shape,rotate=0]
\usetikzlibrary{positioning}
	\tikzstyle{every path}=[line width=2pt]
	\node[anchor=south west,inner sep=0] (image) at (0,0) {
\includegraphics[width=\linewidth,page=2,clip,trim={0 0 0 0}]{distribution_of_docking_results_v2_left.pdf}};
\begin{scope}[x={(image.south east)},y={(image.north west)}]
\draw node[shape=rectangle,minimum size=0.3cm,anchor=center,rotate=0,fill=white] (c) at (0.1,0.9) {(e)};
\draw node[shape=rectangle,minimum size=0.3cm,anchor=center,rotate=0,fill=white] (c) at (0.1,0.5) {(f)};
 	\end{scope}
\end{tikzpicture}

\phantomsubcaption
\label{figsub:vina_results_distribution_c}
\end{subfigure}

\begin{subfigure}[t]{\linewidth}
\phantomsubcaption
\label{figsub:vina_results_distribution_d}
\end{subfigure}
\caption{(a) The names of the segments along the heterodimer. Segments HEAD, TAIL and the linkers L1, L12 are unstructured, while segments 1A, 1B and 2 are coiled coils. (b) Positions of charged residues in the keratin heterodimer K35/K85. Acidic and basic residues are colored red and blue, respectively. Locations of the predicted top poses obtained via AutoDock VINA on the full heterodimer for the ligands DIC (c) and PFBS (d), colored by their VINA score. Number of poses colored by the VINA score along the long heterodimer axis $z$ with $\Delta z=4.94$~\si{\angstrom} for DIC (e) and PFBS (f).}
\label{fig:itwm_result_docking_VINA_otherstyle}
\end{figure*}
In this section, the equilibration of the heterodimer is discussed and the result obtained from docking two ligands is presented. High affinity sites are then subjected to long--time molecular dynamics simulation. Stable sites are identified and further processed in multiple short simulations to analyze the interaction energies and molecular interaction fingerprints in detail.

\subsection{K35/K85 heterodimer equilibration}

This study was initiated using the protein heterodimer structure K35/K85 from ref~\citenum{Chou.2012}, modified in this study as described in the methods part, \cref{sec:methods-dimer}.
This structure was relaxed %
in water for $6.6$~\si{\nano\second} to correct for chirality errors and peptide bonds, while keeping the overall elongated shape. The relaxed heterodimer structure is made freely available\cite{K35-K85-molmod}.

To investigate the evolution of the heterodimer structure, it was equilibrated using the protocol defined in \cref{subsec:equiprotocol} over $50$~\si{\nano\second}. During this simulation run, the protein heterodimer did not remain completely elongated as a filament but started to bend. This behavior was previously observed in the literature in a modeling study for a solvated epithelial keratine heterodimer build from K1/K10 by \citeauthor{Bray.2015} in ref \citenum{Bray.2015} and was experimentally found for a similar filament protein vimentin.\cite{Guharoy.2013}

In the following application of docking on the full heterodimer, the elongated model from the $6.6$~\si{\nano\second} equilibration was used, since inside a human hair the heterodimer is expected to be geometrically closer to the situation in an elongated filament even if in contact with water compared to a fully solvated heterodimer.

\subsection{VINA scores of DIC and PFBS on the heterodimer}

In a first step, blind docking of PFBS and DIC anions, see \cref{fig:molecule_structures}, was applied %
to check the existence and variability of binding sites along the structure. A box of size $32\times32\times25$~\si{\cubic\angstrom} was used as the standard search volume and was shifted by $20$~\si{\angstrom} in each dimension so that each residue is contained at least once. In total $610$ boxes were generated.

The VINA results for both ligands are displayed in \cref{figsub:vina_results_distribution_image_a,figsub:vina_results_distribution_image_b} and show weaker and stronger sites along the heterodimer. In \cref{figsub:vina_results_distribution_c,figsub:vina_results_distribution_d} the distribution of the predicted score is further visualized.  Interpretation regarding real adsorption energies is limited as noted above, however, the scores can be used for a qualitative ranking of the poses. Both ligands exhibit higher affinity (lower scores are better) at the globular HEAD and TAIL segments of the keratine heterodimer. 
Along the helical parts, the surface density of possible high affinity binding sites is reduced compared to that of the HEAD and TAIL segments. Adsorption on the long helical segments is assumed as well, especially in an higher order assemble of two or more dimers. However, as the globular end segments fold back around the helical elements, in principle more buried sites are to be expected.
The best score values are --$6.7$ and --$6.5$ for DIC and PFBS, respectively. Overall, the score distribution is slightly better for DIC.

In a second step of docking investigations, the focus is on the N-terminal of the heterodimer consisting of the HEAD and 1A segment using a higher resolution of the search grid. The HEAD segment is assumed to be less flexible than the TAIL segment, which is the case for a similar filament protein\cite{Hess.2013} and was also found for an epithelial heterodimer of K1/K10\cite{Bray.2015} as well.

\begin{table}[tb]
    \centering
\caption{Properties of the systems D and P that are subjected to long time molecular dynamics simulation for the ligands DIC and PFBS, respectively.}
\label{tab:system_sizes_longtimesims}
\resizebox{\columnwidth}{!}{%
\begin{tabular}{@{}c|c|c@{}}
\toprule
system      & D             & P             \\ \midrule
residue numbers & 1-164/456-644 & 1-140/456-620 \\
ligand      & DIC           & PFBS          \\
\# ligands          & 87            & 67            \\
box size & $91\times188\times96$~\si{\cubic\angstrom} & $140\times140\times140$~\si{\cubic\angstrom} \\ 
\# \ch{H2O} molecules & 52995 & 91600\\
salt type & \ch{NaCl} & \ch{KCl}\\ \bottomrule
\end{tabular}%
}
\end{table}

The high stability of the secondary structure in proteins such as $\alpha$--helical elements allows a subdivision of the model to minimize simulation boxes, save computer time, and
prepare a molecular system that is used in \cref{sec:methods-MD} below.

The heterodimer is taken from the $50$~\si{\nano\second} equilibrated structure and reduced to the residue range defined in \cref{tab:system_sizes_longtimesims}.  For easier differentiation, the selected residue ranges will be called HEAD region.
The truncated heterodimer was solvated in water, neutralized and subjected to equilibration in the NpT ensemble following the protocol of \cref{subsec:equiprotocol} for $220$~\si{\nano\second} in total.

Docking analysis was conducted on the segmented protein structures again with the same strategy as before but with a higher resolution. The search space volume was $20^3$~\si{\cubic\angstrom} and shifted in all three dimensions by $10$~\si{\angstrom} to fully sample the segment. This generated $\approx 1650$ boxes. Neither qualitative nor quantitative differences were found compared to the first application,
leading to the conclusion that the truncated structures are sufficient to study the binding of ligands at the HEAD region.

In order to prepare input structures for the following molecular dynamics simulations, which should have a higher predictive power than docking, 
around $500$ poses from the second wave of docking with scores between --$6.8$ and --$0.18$ were collected and filtered, favoring poses with a lower score and removing
geometrically very similar results. Some poses with VINA scores higher than --$3$ were kept to allow a potential identification in the full molecular dynamics simulation of stable poses not found by docking. In this way, up to 87 ligands were selected for further simulation in the full molecular dynamics production runs described in the next section.

\subsection{Molecular dynamics modeling of binding}\label{sec:methods-MD}

\begin{figure}[tb]
    \begin{subfigure}[]{\linewidth}
		\caption{}
  \label{subfig:snapshots_DIC}

\begin{tikzpicture}[transform shape,rotate=0]
    \usetikzlibrary{positioning}
    \tikzstyle{every path}=[line width=2pt]
    \node[anchor=south west,inner sep=0] (image) at (0,0) {
    \includegraphics[width=\linewidth]{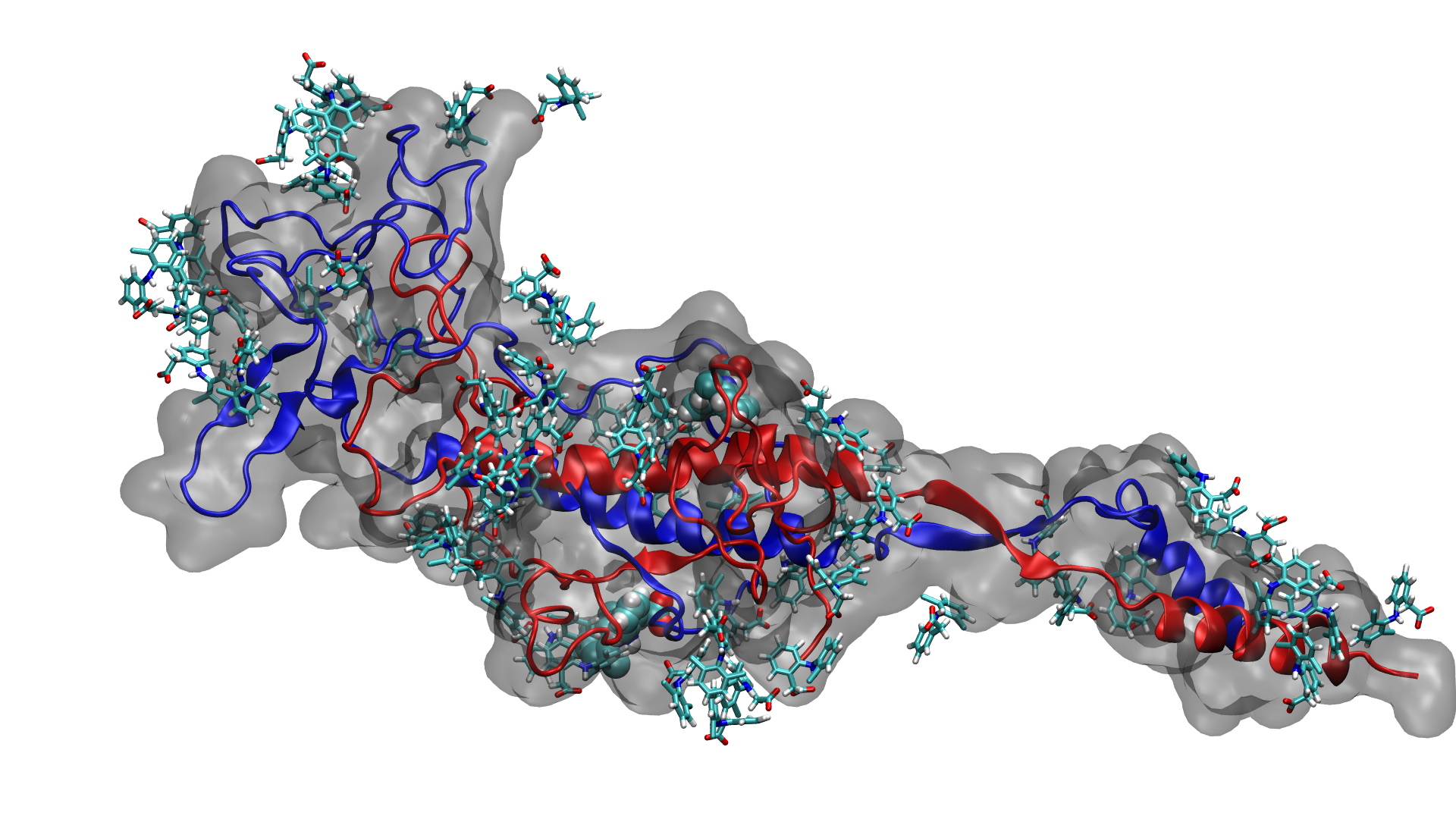}};
    \begin{scope}[x={(image.south east)},y={(image.north west)}]

        \draw node[shape=circle,draw,fill=none,color=black,minimum size=1cm] (A) at (0.5,0.5) {};
		\draw node[shape=rectangle,minimum size=0.15cm,inner sep=0,outer sep=0,anchor=east,rotate=0,fill=none,above right=of A] (B)  {\footnotesize DO23};
		\draw [-] (A) -- (B);
        \draw node[shape=circle,draw,fill=none,color=black,minimum size=1cm] (C) at (0.42,0.25) {};
		\draw node[shape=rectangle,minimum size=0.15cm,inner sep=0,outer sep=0,anchor=east,rotate=0,fill=none] (D) at ($(C) + (-0.1,-0.15)$) {\footnotesize DO61};%
		\draw [-] (C) -- (D);
\end{scope}

\end{tikzpicture}
  
	\end{subfigure}
 \vspace*{-1cm}
 	\begin{subfigure}[]{\linewidth}
        \caption{}
          \label{subfig:snapshots_PFBS}

\begin{tikzpicture}[transform shape,rotate=0]
    \usetikzlibrary{positioning}
    \tikzstyle{every path}=[line width=2pt]
    \node[anchor=south west,inner sep=0] (image) at (0,0) {
    \includegraphics[width=\linewidth,clip,trim={0 0 250 200}]{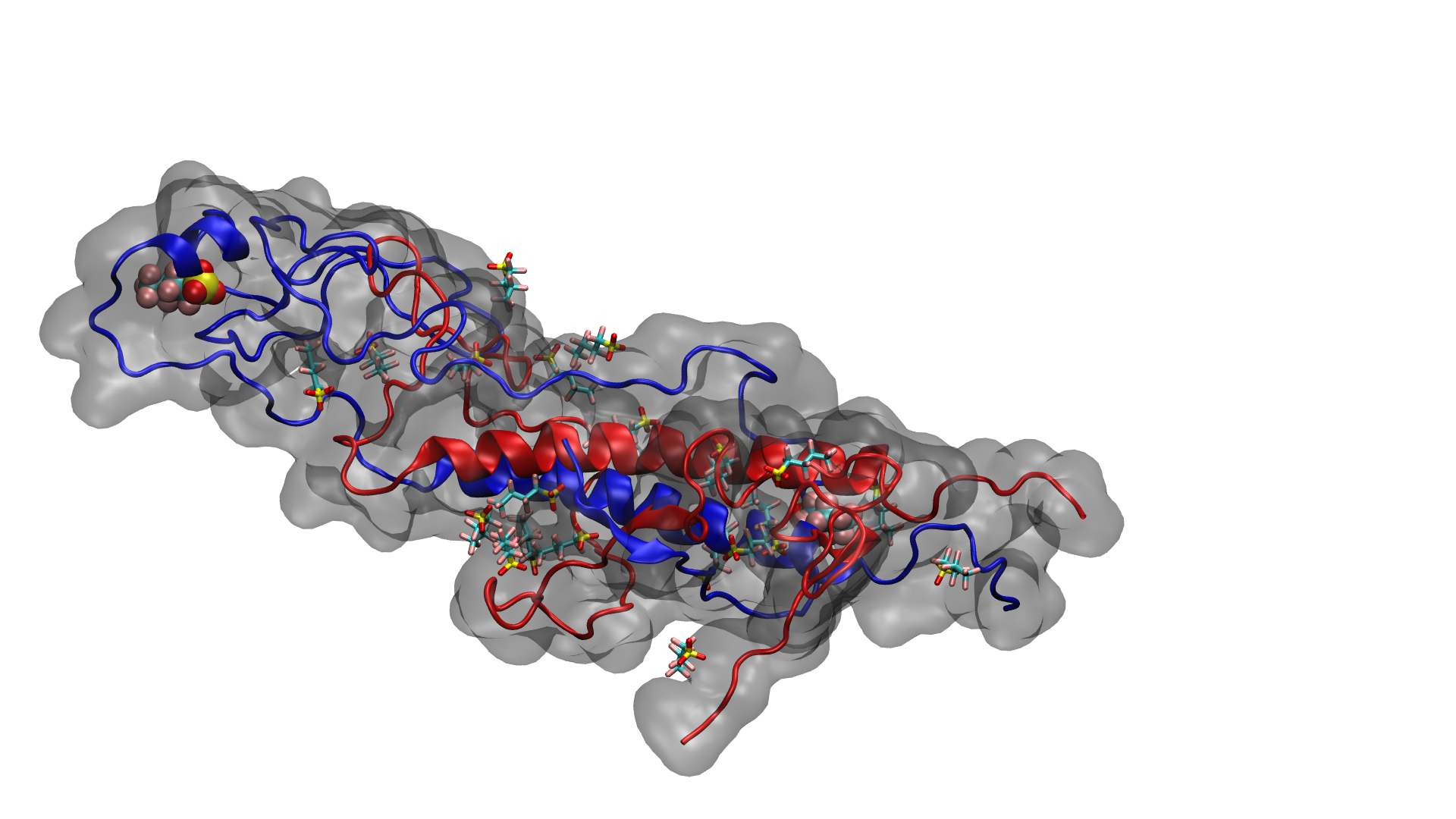}};
    \begin{scope}[x={(image.south east)},y={(image.north west)}]

        \draw node[shape=circle,draw,fill=none,color=black,minimum size=1cm] (A) at (0.15,0.8) {};
		\draw node[shape=rectangle,minimum size=0.15cm,inner sep=0,outer sep=0,anchor=east,rotate=0,fill=none,below=of A] (B)  {\footnotesize PO64};
		\draw [-] (A) -- (B);
        \draw node[shape=circle,draw,fill=none,color=black,minimum size=1cm] (C) at (0.65,0.45) {};
		\draw node[shape=rectangle,minimum size=0.15cm,inner sep=0,outer sep=0,anchor=east,rotate=0,fill=none,above right=of C] (D)  {\footnotesize PO36};
		\draw [-] (C) -- (D);
\end{scope}

\end{tikzpicture}

	\end{subfigure}
 \vspace*{0.75cm}
	\caption{Snapshots from $1.25$~\si{\micro\second} trajectories. The secondary structure of K35 and K85 are colored in red and blue, respectively. The full protein surface is shown in gray with partial transparency. Only ligands close to the protein are shown for clarity. (a) System D: DIC molecules close to the protein after $0.17$~\si{\micro\second} and (b) system P with PFBS after $1.2$~\si{\micro\second}. Note that for DIC a slightly larger section of the protein was used, i.e. the short helical part on the right corresponds to the 1B segment (a), see \cref{tab:system_sizes_longtimesims}. The positions of the two main sites are highlighted by circles, respectively.}
 \label{fig:snapshots_longtime}
\end{figure}
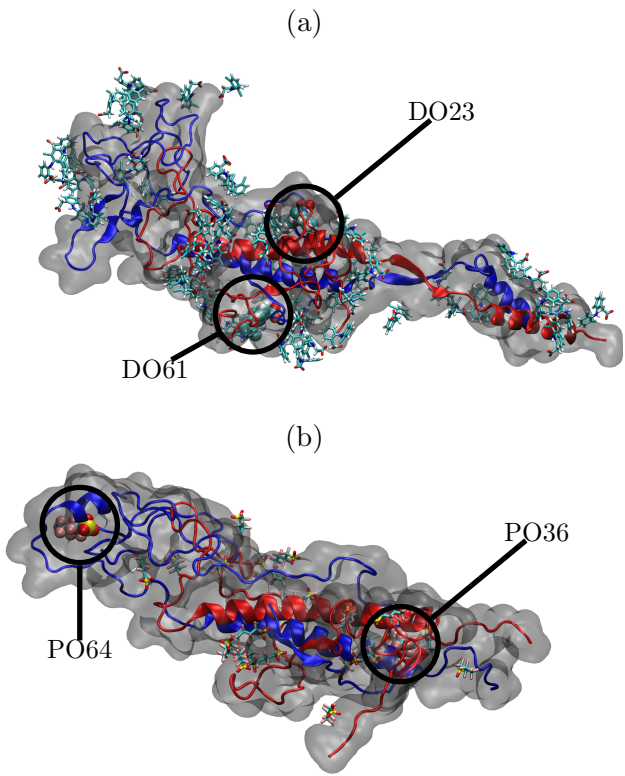

The results described in this section are obtained using the GPU resident mode of NAMD3alpha9.

To verify and evaluate the prediction of sites from AutoDock VINA, the interaction between the ligands and the HEAD region of the protein heterodimer was investigated using MD.
In the following, two molecular systems D and P are introduced for the two ligands DIC and PFBS, respectively.

The system D was constructed from the HEAD region and the previously identified poses of DIC, dissolved again in water, and neutralized at an \ch{NaCl} ion concentration of $0.15$~\si{\mol\per\litre}. It was subjected to the standard equilibration protocol as described in \cref{subsec:equiprotocol} for another $2$~\si{\nano\second}.
Subsequently, a simulation of $1.25$~\si{\micro\second} was performed in the NpT ensemble. 
The evolution of system D revealed a high mobility of the last 20 residues that are part of segment 1B, see \cref{fig:itwm_result_docking_VINA_otherstyle}.

Therefore, for the second system P, the section was truncated at the unstructured linker L1 instead of the helical part, see \cref{tab:system_sizes_longtimesims} for the residue number ranges. $67$ poses for PFBS were then selected from the docking output on the reduced structure. %
The final configurations for system D and P are described in \cref{tab:system_sizes_longtimesims} and snapshots of the trajectories are shown in \cref{fig:snapshots_longtime}.

Throughout the two long trajectories, stable sites, dissociation as well as association processes could be observed. Only some of the poses suggested by the docking turned out to be stable, sometimes owing to the flexibility of the unstructured part of the HEAD segment. 

\begin{figure}[b]
    \centering
\includegraphics{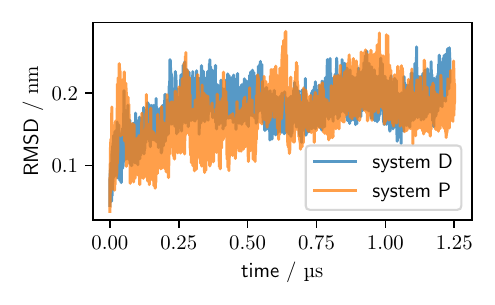}
    \caption{Evolution of the RSMD of the two longtime simulations of segment 1A for sytem D and system P.}
    \label{fig:rmsd_longtime_sims}
\end{figure}

The backbone stability of the helical segment 1A was evaluated by root--mean--square deviation (RMSD) and is shown in \cref{fig:rmsd_longtime_sims}, exhibiting only a small change on a scale of $2$~\si{\angstrom} over a simulation time of more than $1$~\si{\micro\second}.

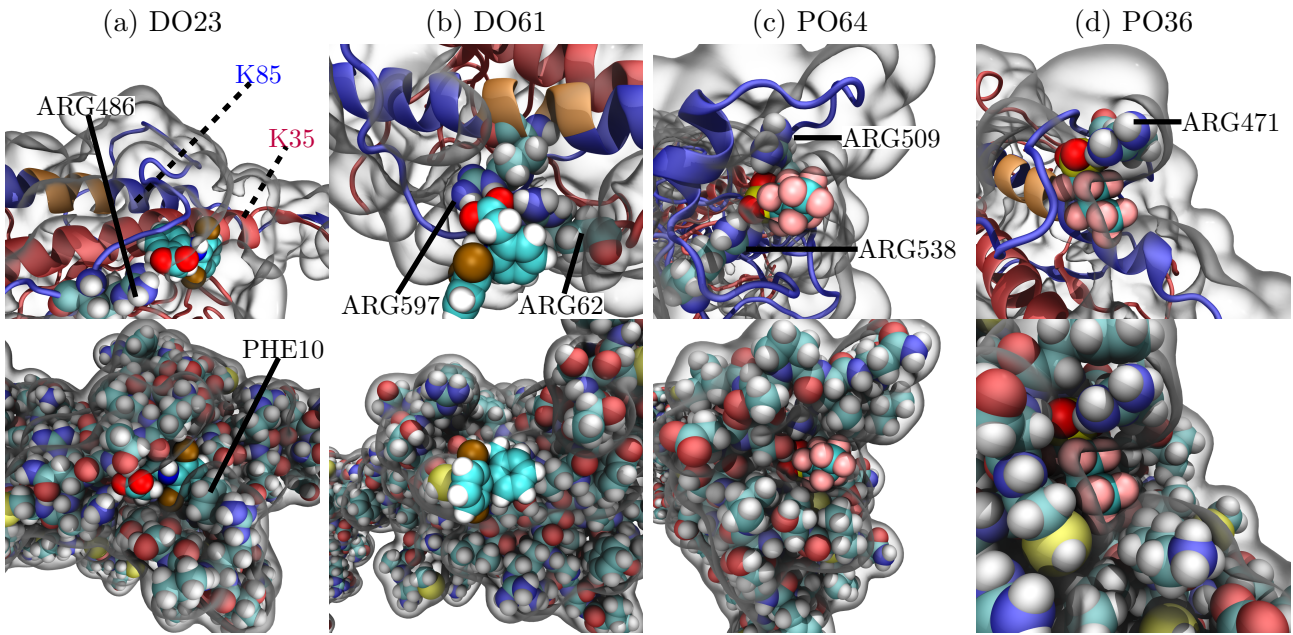
\begin{figure*}[tb]
\centering
\begin{subfigure}[t]{0.235\textwidth}
\caption{DO23}
\label{fig:bindsite_dic_DO23_img2}

		\begin{tikzpicture}[transform shape,rotate=0]
	\usetikzlibrary{positioning}
	\tikzstyle{every path}=[line width=2pt]
	\node[anchor=south west,inner sep=0] (image) at (0,0) {
		\includegraphics[width=\textwidth,angle=-90,scale=1.0,trim={200px 250px 600px 500px},clip]{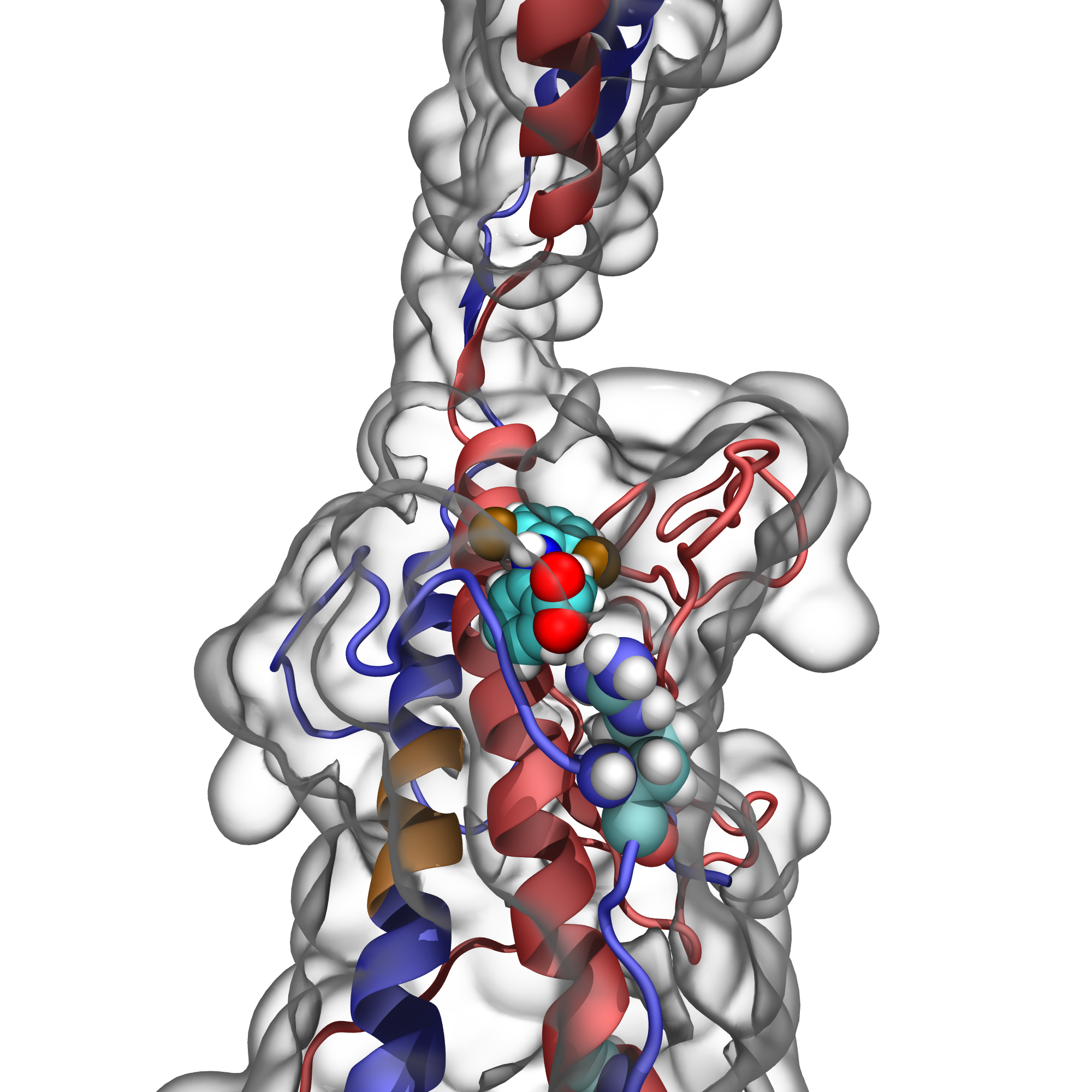}};
	
	\begin{scope}[x={(image.south east)},y={(image.north west)}]
		\draw node[shape=rectangle,minimum size=0.15cm,inner sep=0,outer sep=0,anchor=west,rotate=0,fill=white] (A) at (0.1,0.8) {\footnotesize ARG486};
		\draw [-] (0.4,0.2) -- (A);

        \draw node[shape=rectangle,minimum size=0.15cm,inner sep=0,outer sep=0,anchor=west,rotate=0,fill=white] (K35) at (0.8,0.7) {\footnotesize \color{purple}{K35}};
				\draw [-,dashed,color=black] (0.725,0.45) -- (K35);
								\draw node[shape=rectangle,minimum size=0.15cm,inner sep=0,outer sep=0,anchor=west,rotate=0,fill=white] (K85) at (0.7,0.9) {\footnotesize \color{blue}{K85}};
				\draw [-,dashed] (0.4,0.5) -- (K85);

	\end{scope}
	
\end{tikzpicture}

\end{subfigure}
\begin{subfigure}[t]{0.235\textwidth}
\caption{DO61}
\label{fig:bindsite_dic_DO61_img2}

	\begin{tikzpicture}[transform shape,rotate=0]
	\usetikzlibrary{positioning}
	\tikzstyle{every path}=[line width=2pt]
	\node[anchor=south west,inner sep=0] (image) at (0,0) {
		\includegraphics[width=\textwidth,angle=0,scale=1.0,trim={500px 750px 500px 250px},clip]{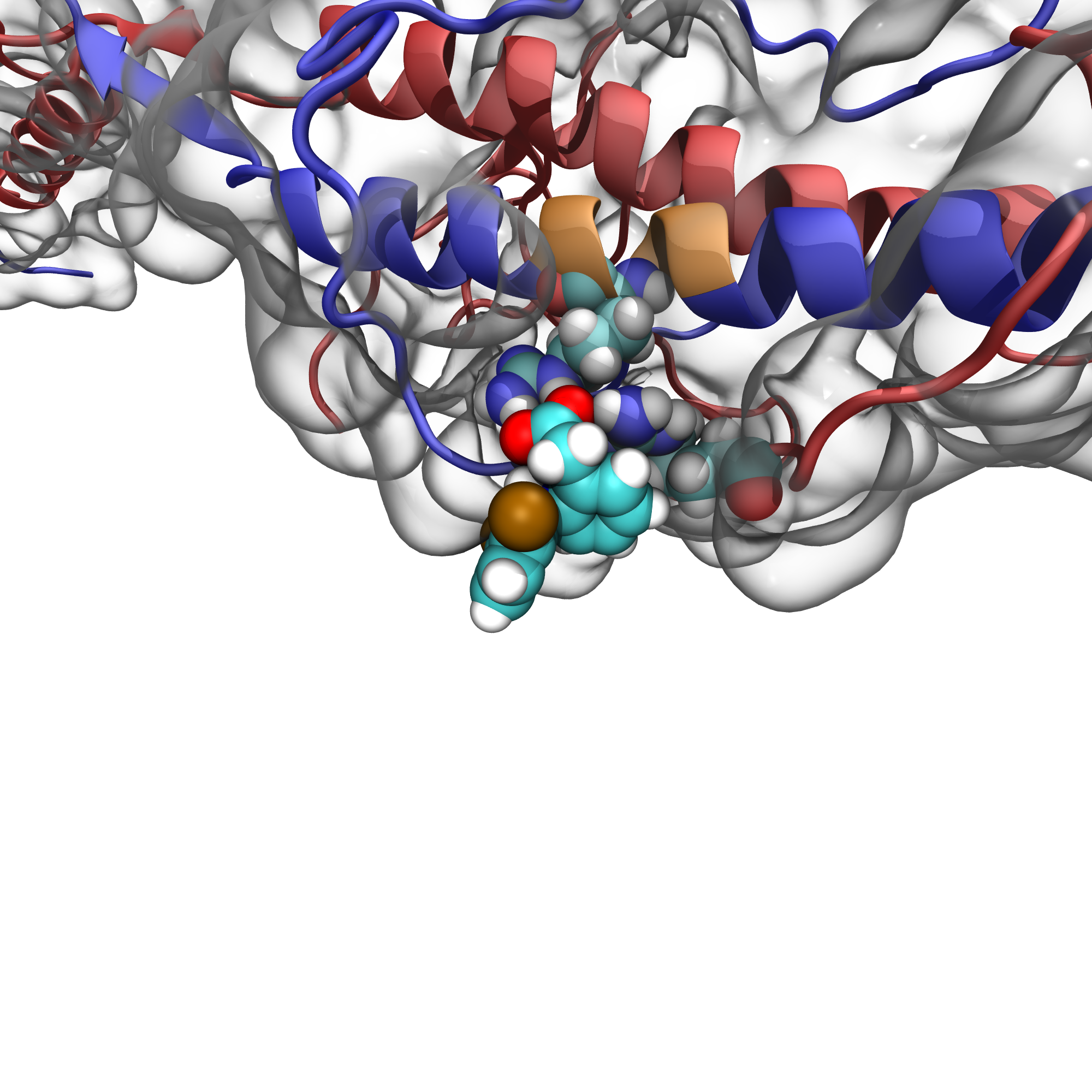}};
	
	\begin{scope}[x={(image.south east)},y={(image.north west)}]
		\draw node[shape=rectangle,minimum size=0.3cm,anchor=center,inner sep=0,outer sep=0,rotate=0,fill=none] (A) at (0.2,0.175) {\footnotesize ARG597};
		\draw [-] (0.4,0.5) -- (A);
		\draw node[shape=rectangle,minimum size=0.3cm,inner sep=0,outer sep=0,anchor=west,rotate=0,fill=white] (B) at (0.6,0.175) {\footnotesize ARG62};
		\draw [-] (0.8,0.4) -- (B);

	\end{scope}
	
\end{tikzpicture}

\end{subfigure}
\begin{subfigure}[t]{0.235\textwidth}
\caption{PO64}
\label{fig:bindsite_pfbs_PO64_img2}

		\begin{tikzpicture}[transform shape,rotate=0]
	\usetikzlibrary{positioning}
	\tikzstyle{every path}=[line width=2pt]
	\node[anchor=south west,inner sep=0] (image) at (0,0) {
		\includegraphics[width=\textwidth,angle=0,scale=1.0,trim={500px 500px 500px 500px},clip]{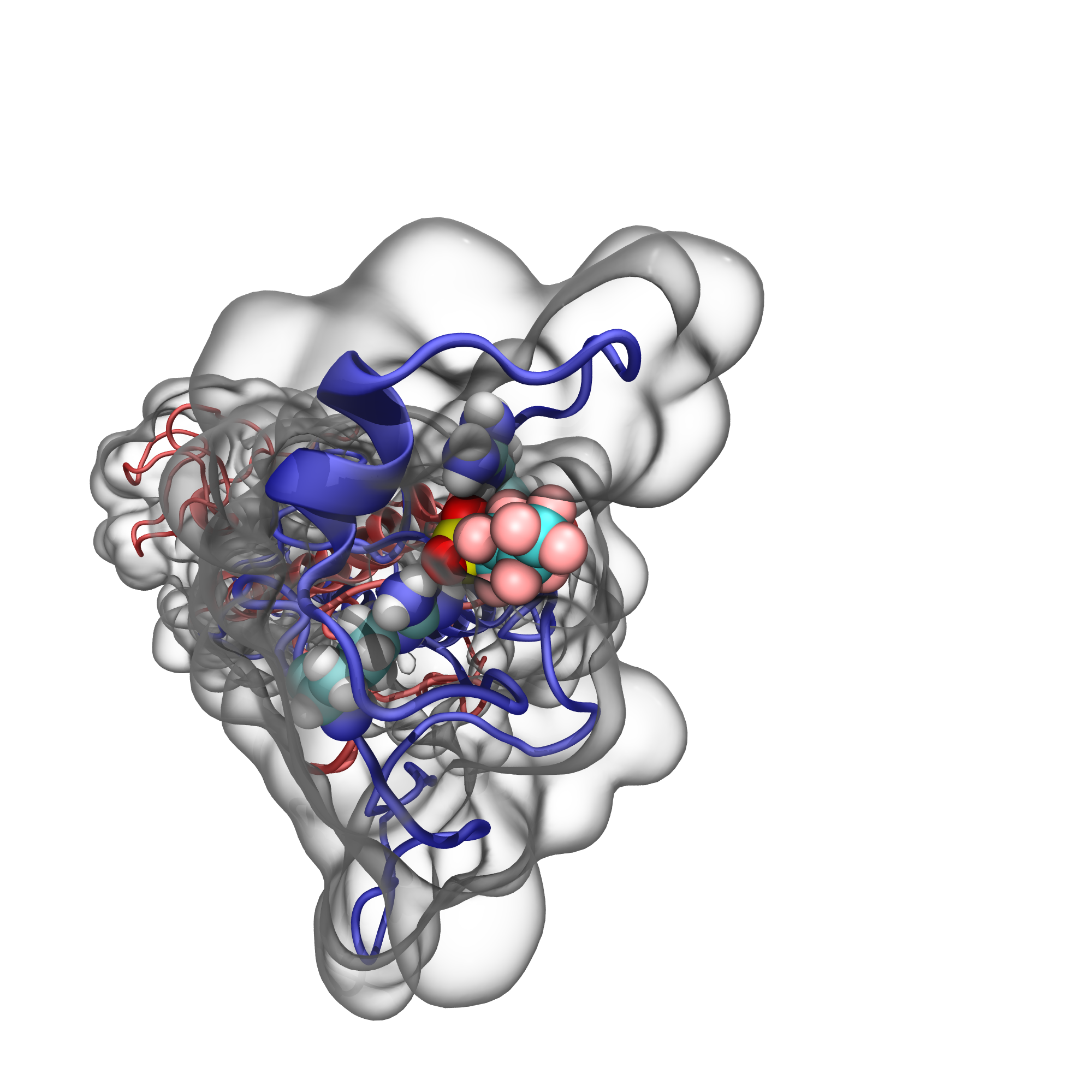}};
	
	\begin{scope}[x={(image.south east)},y={(image.north west)}]
		\draw node[shape=rectangle,minimum size=0.3cm,inner sep=0,outer sep=0,anchor=west,rotate=0,fill=white] (A) at (0.6,0.7) {\footnotesize ARG509};
		\draw [-] (0.45,0.7) -- (A);
		\draw node[shape=rectangle,minimum size=0.3cm,inner sep=0,outer sep=0,anchor=west,rotate=0,fill=none] (B) at (0.65,0.35) {\footnotesize ARG538};
		\draw [-] (0.3,0.35) -- (B);

	\end{scope}
	
\end{tikzpicture}

\end{subfigure}
\begin{subfigure}[t]{0.235\textwidth}
\caption{PO36}
\label{fig:bindsite_pfbs_P36_img2}

		\begin{tikzpicture}[transform shape,rotate=0]
	\usetikzlibrary{positioning}
	\tikzstyle{every path}=[line width=2pt]
	\node[anchor=south west,inner sep=0] (image) at (0,0) {
		\includegraphics[width=\textwidth,angle=0,scale=1.0,trim={250px 250px 250px 250px},clip]{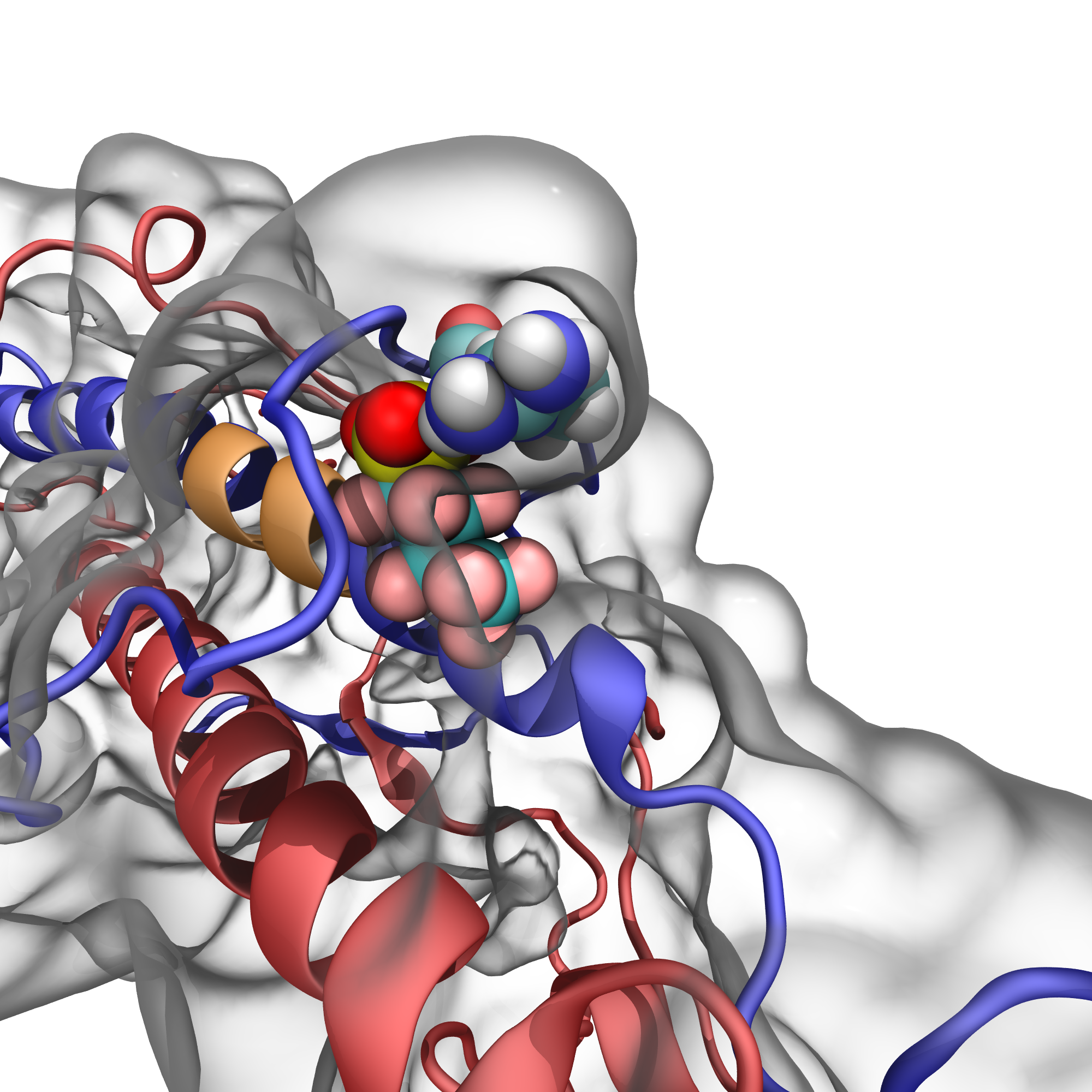}};
	
	\begin{scope}[x={(image.south east)},y={(image.north west)}]
		\draw node[shape=rectangle,minimum size=0.3cm,inner sep=0,outer sep=0,anchor=west,rotate=0,fill=none] (A) at (0.65,0.75) {\footnotesize ARG471};
		\draw [-] (0.5,0.75) -- (A);

	\end{scope}
	
\end{tikzpicture}

\end{subfigure}\\[-3ex]
\begin{subfigure}[t]{0.235\textwidth}

		\begin{tikzpicture}[transform shape,rotate=0]
	\usetikzlibrary{positioning}
	\tikzstyle{every path}=[line width=2pt]
	\node[anchor=south west,inner sep=0] (image) at (0,0) {
		\includegraphics[width=\textwidth,angle=-90,scale=1.0,trim={400px 400px 400px 400px},clip]{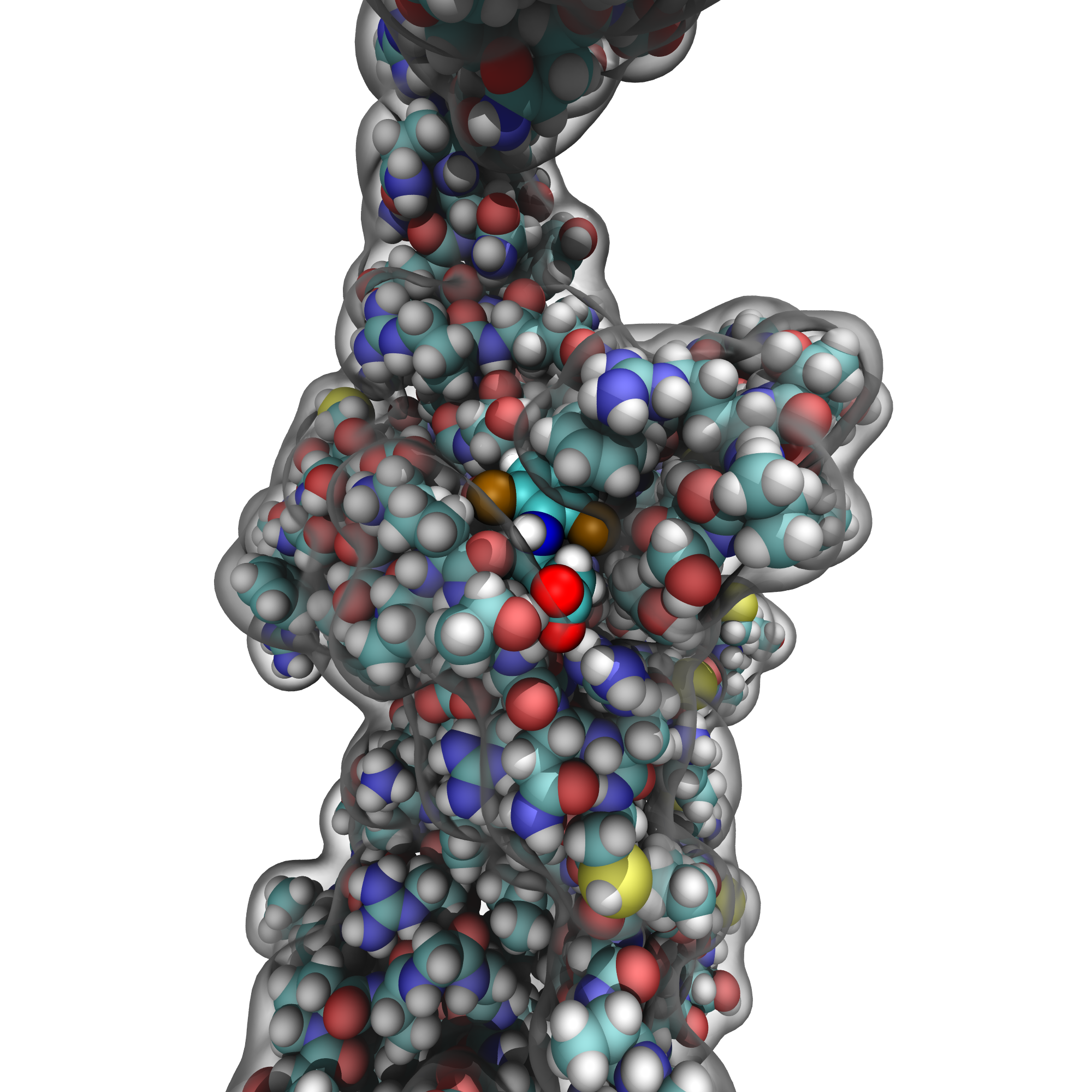}};
	
	\begin{scope}[x={(image.south east)},y={(image.north west)}]
		\draw node[shape=rectangle,minimum size=0.15cm,inner sep=0,outer sep=0,anchor=west,rotate=0,fill=white] (A) at (0.75,0.9) {\footnotesize PHE10};
		\draw [-] (0.65,0.45) -- (A);

	\end{scope}
	
\end{tikzpicture}

\phantomsubcaption
\label{fig:bindsite_dic_DO23_img1}
\end{subfigure}
\begin{subfigure}[t]{0.235\textwidth}
\includegraphics[width=\textwidth,angle=0,scale=1.0,trim={400px 400px 400px 400px},clip]{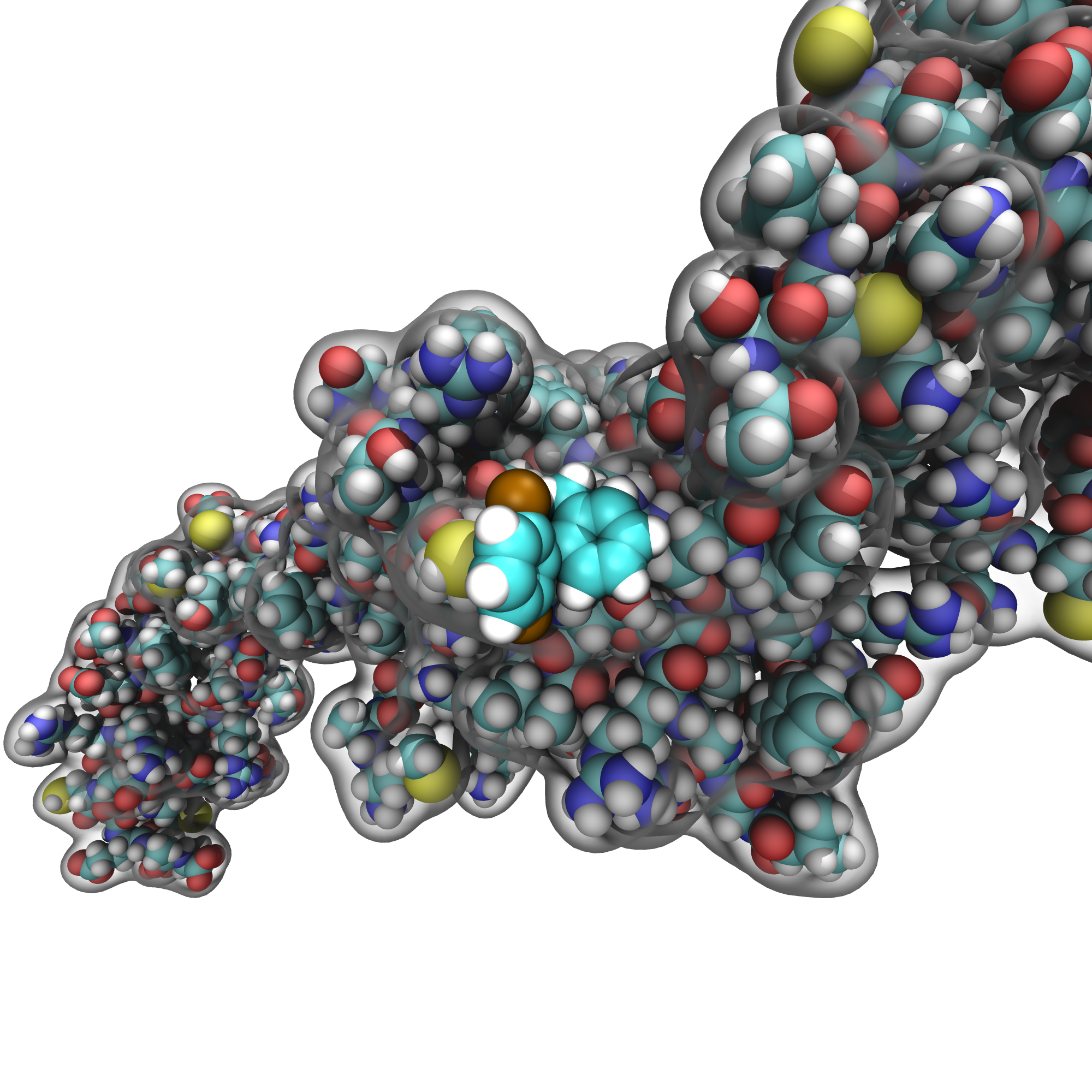}
\label{fig:bindsite_dic_DO61_img1}
\end{subfigure}
\begin{subfigure}[t]{0.235\textwidth}
\includegraphics[width=\textwidth,angle=0,scale=1.0,trim={300px 350px 500px 450px},clip]{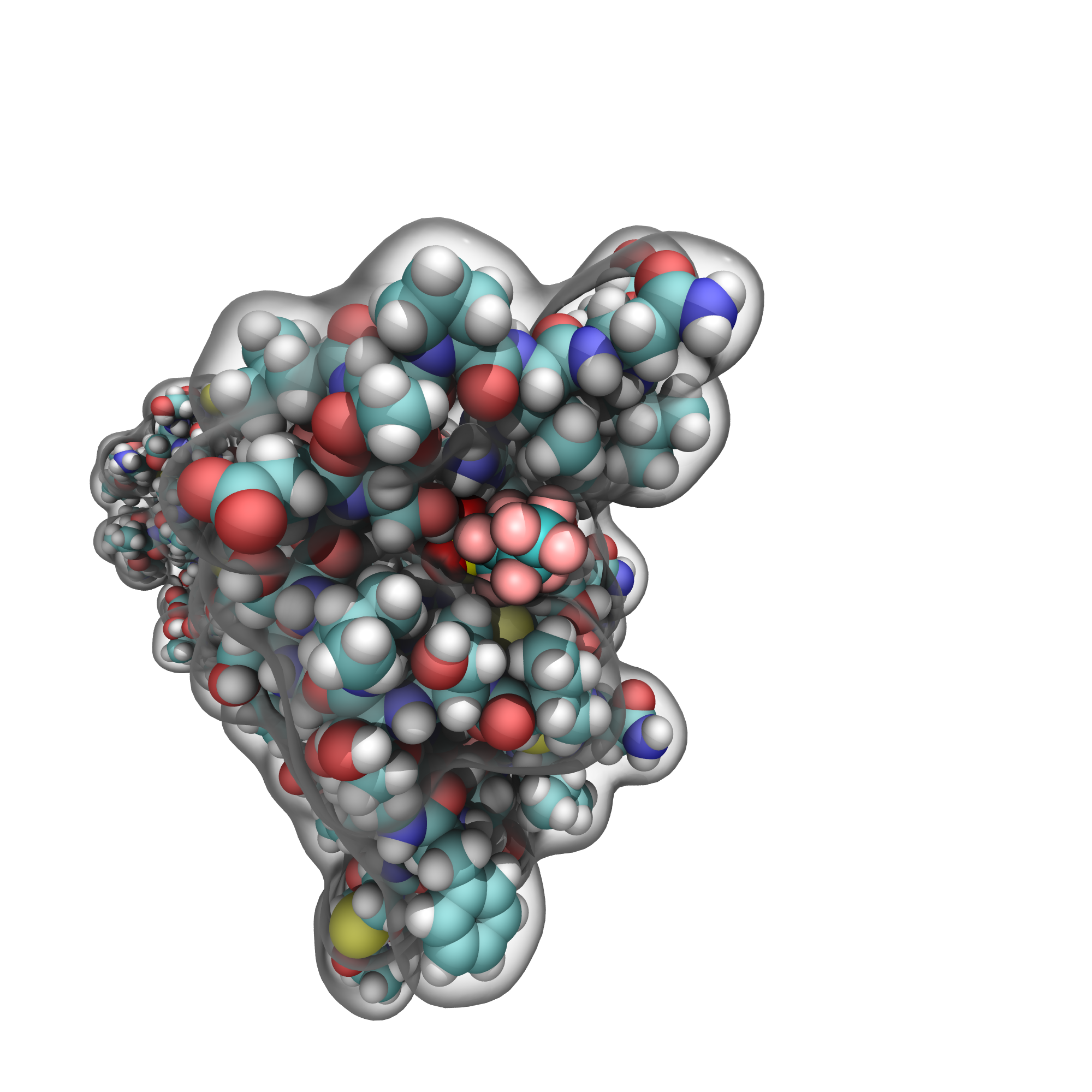}
\label{fig:bindsite_pfbs_PO64_img1}
\end{subfigure}
\begin{subfigure}[t]{0.235\textwidth}
\includegraphics[width=\textwidth,angle=0,scale=1.0,trim={400px 400px 400px 400px},clip]{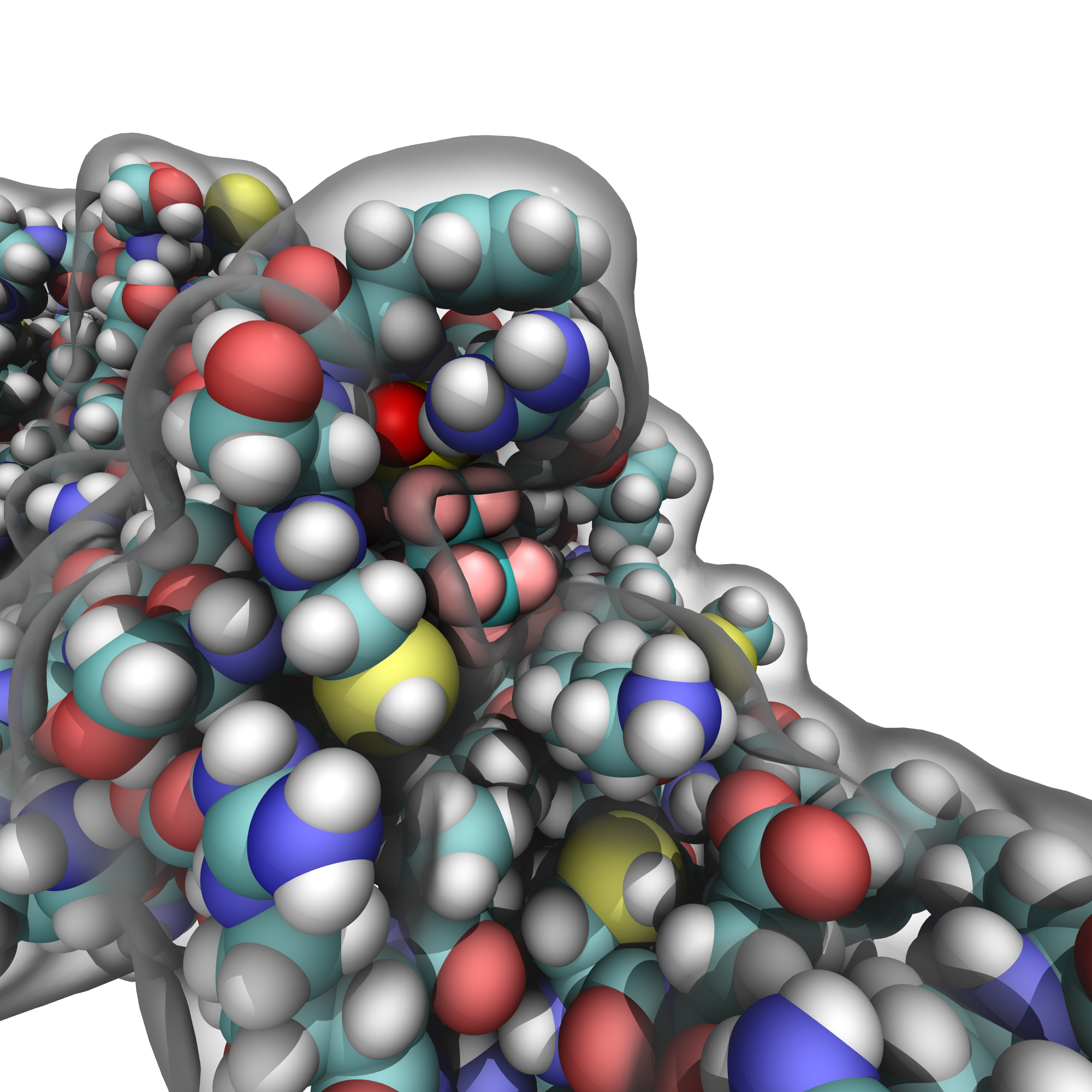}
\label{fig:bindsite_pfbs_PO36_img1}
\end{subfigure}
\caption{Top row: Zoomed views (a-d) of the binding site with the basic residues shown to highlight the electrostatic interaction partners of the ligands. Bottom row: A wider view of the location of binding sites with highest affinity determined by the LIE method at the HEAD region. The ligand is shown in brighter colors than the protein residues. Atom colors: carbon (cyan), hydrogen (white), oxygen (red), nitrogen (blue), sulfur (yellow), chlorine (brown) and fluorine (pink). The secondary structure of the protein heterodimer is shown in the upper panel in red and blue for K35 and K85 (see (a) top row), respectively. The protein shell is shown as a transparent QuickSurf representation. As an anchoring point the backbone of K85 was additionally colored in orange for the residue number range 595 to 600 in the top row for a, b and d. The locations are highlighted in \cref{fig:snapshots_longtime} as well.}
\label{fig:bindingsites_postequi}
\end{figure*}
\begin{figure*}[tb]
\begin{subfigure}[t]{0.24\textwidth}
\includegraphics[width=1.75in]{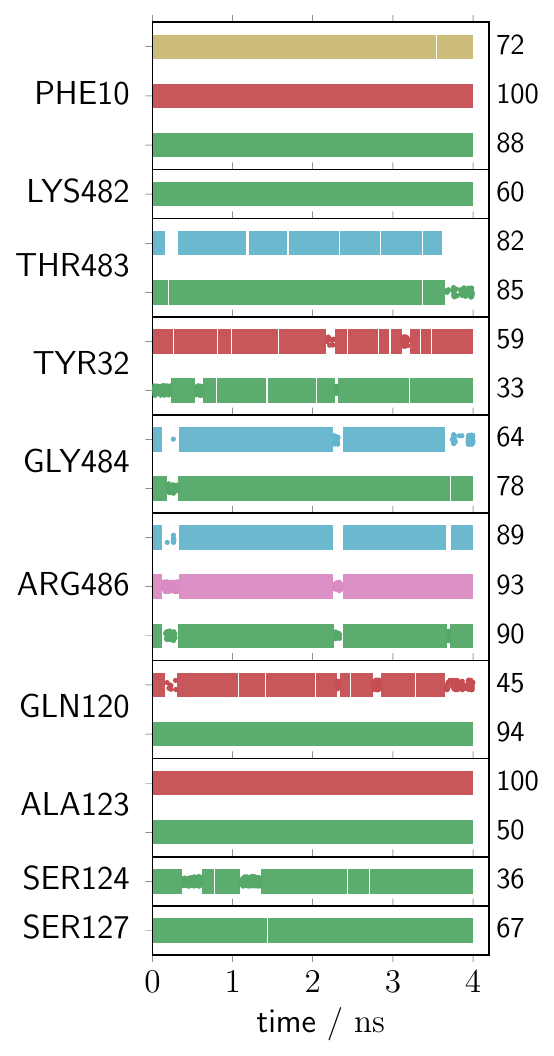}
\caption{DO23}
\label{fig:timelineDO23}
\end{subfigure}
\begin{subfigure}[t]{0.24\textwidth}
\includegraphics[width=1.75in]{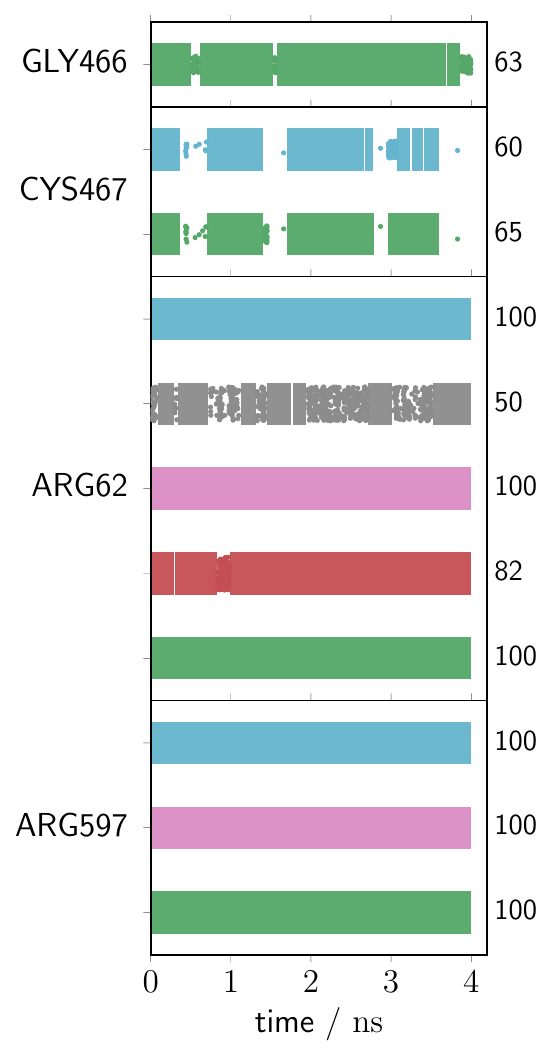}
\caption{DO61}
\label{fig:timelineDO61}
\end{subfigure}
\begin{subfigure}[t]{0.24\textwidth}
\includegraphics[width=1.75in]{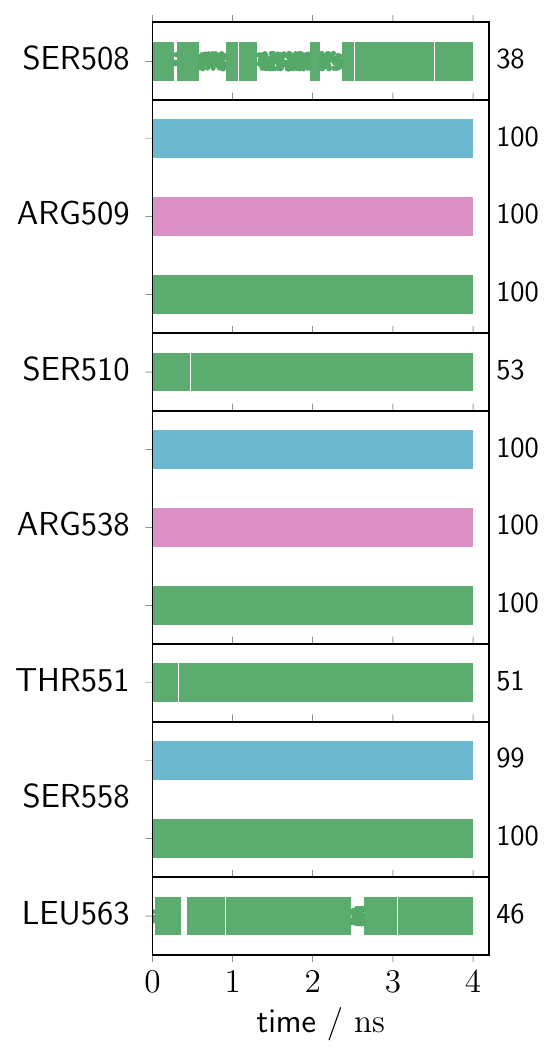}
\caption{PO64}
\label{fig:timelinePO64}
\end{subfigure}
\begin{subfigure}[t]{0.24\textwidth}
\includegraphics[width=1.75in]{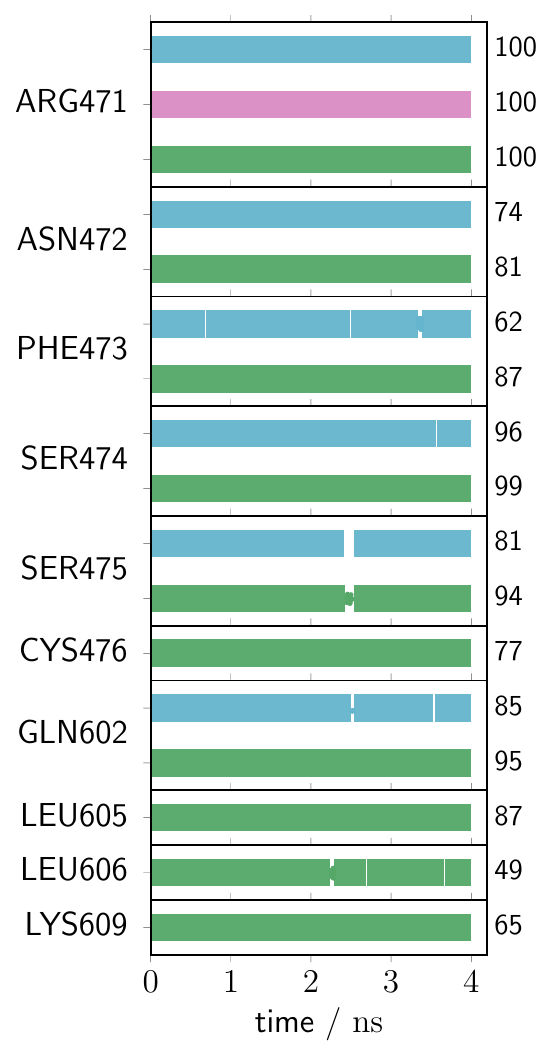}
\caption{PO36}
\label{fig:timelinePO36}
\end{subfigure}
\vspace*{-5mm}
\begin{center}
\begin{subfigure}[t]{\linewidth}
\includegraphics[width=\linewidth]{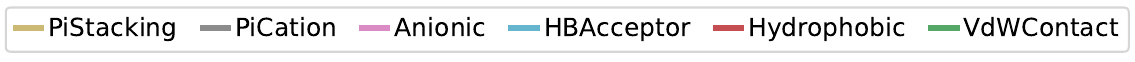}
\end{subfigure}
\end{center}
\vspace*{-5mm}

\caption{Timeline of relevant interactions of a $4$~\si{\nano\second} simulation between DIC (a,b) and PFBS (c,d) ligands and residues of the HEAD region of the keratin heterodimer used to determine the free energy of adsorption. {The number on the right represents the overall stability of an interaction in percent. The trajectory output period was $0.5$~\si{\pico\second}.} Interactions that occur almost continuously are visualized as a block; the remaining data are shown as jittered points. {Points are joined into a block if its spans at least $50$~\si{\pico\second} with interruptions of less than $10$~\si{\pico\second} allowed.}}
\label{fig:itwm_result_interaction_timelines_short}
\end{figure*}

Then, stable poses were defined from the interaction sites that span at least $200$~\si{\nano\second} in time and are sorted accordingly. The region of interest consists of the HEAD and the 1A segment of the heterodimer, therefore binding sites at the artificial interface were neglected as well as at the short helical part of segment 1B contained in system D, see \cref{subfig:snapshots_DIC}.

Five ligands were selected from both systems using different snapshots of the long-time simulation trajectories. Then, single-ligand systems were constructed by removing all other ligands and excess ions, leaving a system as shown without the counter ions in \cref{fig:system_illustration}.
The influence of the salt on the binding affinity was investigated but was found to be negligible in this case.
The reduced systems are then subjected to a short minimization and equilibration of $10$~\si{\pico\second} in the NpT ensemble as before. Then, $7$ simulations are performed with $4$~\si{\nano\second} each. Information about the specific types of interaction and the interaction energy between the ligand and the system is gathered. The energy averages for a ligand molecule far from the protein corresponding to $\langle E_{\mathrm{vdW,EL}}^{\mathrm{free}}\rangle$ are calculated from $3$ simulations for both DIC and PFBS, respectively. The free energy $\Delta G$ is then determined from the vdW and the electrostatic interaction energy using the standard LIE model from \cref{eq:model_lie_s}.

\begin{table}[tb]
\caption{Adsorption free energy obtained via the LIE method and interaction energy differences as defined in \cref{eq:model_lie_s}. All values in units of \si{\kcal\per\mol}.}
\label{tab:LIE_values}
\resizebox{\columnwidth}{!}{%
\begin{tabular}{@{}l|rrr@{}}
\toprule
   \multicolumn{1}{c}{ligand}             &               \multicolumn{1}{c}{$\Delta$G} & \multicolumn{1}{c}{$\Delta E_{\mathrm{EL}}$} &  \multicolumn{1}{c}{$\Delta E_{\mathrm{vdW}}$}\tabularnewline
\midrule
DO23      &  --$5.53 \pm 0.31$ &   --$6.39 \pm 0.56$ &  --$12.99 \pm 0.18$             \\
DO61        &  --$5.10 \pm 0.73$ &   --$8.36 \pm 1.30$ &   --$5.13 \pm 0.46$           \\
DO74        &  --$4.52 \pm 1.42$ &   --$7.01 \pm 2.44$ &   --$5.65 \pm 1.11$           \\
DO4         &  --$4.16 \pm 1.69$ &   --$4.50 \pm 3.27$ &  --$10.60 \pm 0.31$           \\
DO70        &  --$1.00 \pm 1.36$ &    $0.17 \pm 2.45$ &   --$6.02 \pm 0.75$            \\
\midrule
PO64            &  --$5.73 \pm 0.83$ &   --$8.85 \pm 1.50$ &   --$7.27 \pm 0.47$       \\
PO36            &  --$3.70 \pm 0.75$ &   --$2.79 \pm 1.35$ &  --$12.83 \pm 0.43$       \\
PO66            &  --$1.60 \pm 1.69$ &    $0.70 \pm 3.20$ &  --$10.86 \pm 0.49$        \\
PO34            &  --$1.16 \pm 1.40$ &    $1.68 \pm 2.59$ &  --$11.13 \pm 0.55$        \\
PO35            &  $-0.85 \pm 0.66$  &    $1.17 \pm 1.21$ &   --$7.99 \pm 0.31$        \\

\bottomrule
\end{tabular}%
}%
\end{table}

The schematics of two interaction sites for DIC and PFBS are illustrated in \cref{fig:bindingsites_postequi}, respectively. For all four sites shown, the electrostatic interaction is the dominant contribution to the free energy $\Delta G$, see \cref{tab:LIE_values} that lists the value of $\Delta G$, next to the vdW and EL contributions. The dominance of the EL interaction on $\Delta G$ also originates in the value of $\beta$ in \cref{eq:model_lie_s}. Therefore, in the upper panel of \cref{fig:bindingsites_postequi} the basic residue (here only arginine) with their positively charged nitrogen-containing side group is highlighted.

The interaction timelines for selected single ligand setups are shown in \cref{fig:itwm_result_interaction_timelines_short} corresponding to the sites in \cref{fig:bindingsites_postequi}. Details regarding the basics of specific interaction are given in the \cref{sec:methods} and will be discussed below. A interaction is stated to be relevant if it is present in at least $30$\% of the trajectory and omitted in visualization otherwise. Overall, the timelines reveal steady interactions with most of the residue in the corresponding pocket. However, some interactions are not persistent during the whole simulation time, but get interrupted and restored again. For instance, the vdW contact between DO23 and residue TYR32 and the hydrogen bond to GLY484, see \cref{fig:bindingsites_postequi}.

\begin{figure*}[tb]
	
	\begin{subfigure}[t]{\textwidth}
 \centering
 \includegraphics[page=1,width=\textwidth]{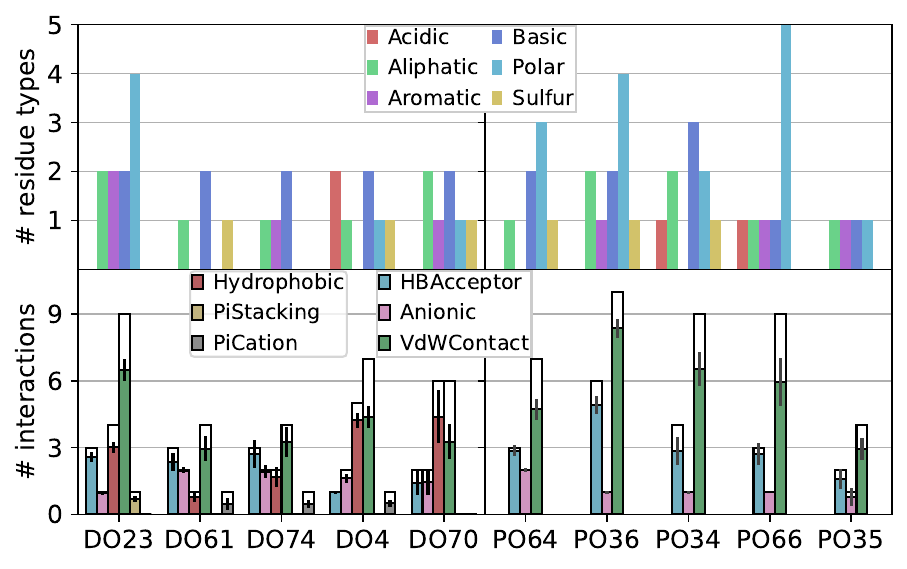}
	\end{subfigure}
	\caption{{The left and right column show the residue and interaction data for DIC and PFBS sites, respectively. Top: Number of residue types in the interaction of the ligands with the heterodimer obtained from a single $4$~\si{\nano\second} trajectory for each site. Multiple contacts with a single residue are only counted once. Bottom row: Stability analysis of interaction types. The full black-framed bar discloses the number of potential interactions of different physical origin between the ligands and the heterodimer. The color-filled part represents the normalized time--averaged occurrence of a given interaction type sampled in $7$ independent runs of $4$~\si{\nano\second}.}}
	\label{fig:residue_type_stats}
\end{figure*}

On the one hand, the interaction of PFBS is dominated by vdW contacts in general, and hydrogen bonds and ionic interaction of the sulfonate group with one or more basic protein residues (arginine or lysine), see \cref{fig:timelinePO64,fig:timelinePO36} and for all sites corresponding to \cref{tab:LIE_values} a summary of residue types present in the right column of \cref{fig:residue_type_stats}.
On the other hand, the ligand DIC participates in one or more hydrophobic interactions, while the major difference between the sites of high and low affinity is the existence of the $\pi$--$\pi$ or $\pi$--cation interaction between DIC and protein residues, respectively. These are rather stable as shown in \cref{fig:itwm_result_interaction_timelines_short,fig:timelineDO23,fig:timelineDO61}, even though polarization effects between cations and aromatic rings are not explicitly accounted for in the force field used, see ref \citenum{Liu.2020}. A higher average stability of the $\pi$--cation interaction of the ligand DIC above $50$\% calculated over the seven trajectories would be expected if those were taken into account, see \cref{fig:residue_type_stats}.

Of the sites under investigation for DIC, four show sufficient binding affinity. The binding site of DO70 is rather loose, cf. \cref{tab:LIE_values}. There, a second anionic interaction occurs over the full trajectory in two runs of $4$~\si{\nano\second}, only partially for another two runs, and does not occur at all for three of the seven runs. This finding emphasizes that multiple sufficiently long simulations are required to properly evaluate a binding site.

For PFBS only two sites could be identified in the HEAD region with an affinity lower than -$3$~\si{\kcal\per\mol} among the $5$ sites investigated.

Furthermore, the contribution of single protein residues was investigated. The electrostatic contributions to the binding site of the DO23 ligand, see \cref{fig:bindsite_dic_DO23_img2,fig:timelineDO23}, originate almost entirely from the interaction with residues ARG486 and THR483. The vdW interactions are mainly driven by the residues GLN120, PHE10, and LYS428 in descending order. 
For the ligand DO61, the contacts with the charged residues ARG597 and ARG62 at this site dominate the overall interaction. The vdW contributions with the residues GLY466, CYS467, and ARG62 are all on the same scale, while for the residue ARG597 the interaction even becomes repulsive.
For both the PO64 and PO36 sites, the major electrostatic interactions result from the basic residues ARG509 and ARG539 for the former and ARG471 for the latter.

Here, the vdW interactions are rather evenly distributed, but are strongest with LEU605 for the PO36 ligand. The same is found for the PO64 ligand, for which the main vdW contacts are ARG509 and SER510.

Next, the interactions are analyzed from a higher perspective regarding the abundance of certain amino acid types as non--covalent contact partners and the stability of the contacts of the five sites per ligand corresponding to the entries in \cref{tab:LIE_values}. %

The upper panel of \cref{fig:residue_type_stats} shows the number of different types of amino acids in contact with the ligand at specific sites from single trajectories. The lower panel of \cref{fig:residue_type_stats} shows the overall stability of the interactions by averaging the seven simulations for each site, respectively. A black-framed bar is partially filled with color when a contact is formed, while the fraction of its absence is represented by the unfilled part.

Multiple findings are extracted in the following.
First, \cref{fig:residue_type_stats} highlights that interaction with a specific type of residue, such as aromats, does not guarantee the existence of $\pi$--$\pi$ or $\pi$--cation interactions. For example, in the case of sites DO74 and DO70, contacts are made with TYR461 and THR100, respectively. However, no $\pi$-stacking occurred. The same is true for PO36, which forms a vdW contact with LYS609, but no anionic interaction with its charged side group was detected above the threshold as defined above.

Secondly, for DO70, the vdW contacts form approximately $50$\% of the time, further highlighting its reduced stability. For all other sites, a higher 
stability is found, but for the $\pi$--cation interactions, which fluctuate at the same level. The latter are expected to increase in strength and occurrence when using optimized Lennard-Jones parameters as discussed above.

Furthermore, the differences between the sites DO23 and DO61 reveal that hydrophobic interactions occur more dominantly for the first one, which is rather buried. This is expected because proteins tend to fold in a way that minimizes the contact of hydrophobic residues with water.

The two binding sites of PFBS, PO64 and PO36, both form anionic contacts and hydrogen bonds with high stability. Compared to the other three sites, the strength of PO64 can be explained by its additional anionic interaction, while PO36 has more hydrogen bonds and vdW contacts. Furthermore, PO34 and PO66 interact with the acidic residues ASP106 and GLU121, respectively, which weaken the electrostatic interaction.

In summary, this analysis shows that investigations using molecular fingerprints supports the interpretation of binding sites in greater detail that are not accessible in the experiment.

\subsection{Comparison to other materials}

In order to assess the binding affinities obtained, the results are compared with available computational and experimental data.
Ref \citenum{BouAbdallah.2016} reports a free energy $\Delta G$ of $\approx $--$6.1$~\si{\kcal\per\mol} in $298$~\si{\kelvin} for the interaction between DIC and human serum albumin, the most abundant protein in blood plasma. Furthermore, the binding affinity of DIC is in the range of --$8.2$ to --$5.9$~\si{\kcal\per\mol} on pristine graphene\cite{Veclani.2022}, alignate/carbon-based films\cite{Shamsudin.2022} or hyper cross-linked polymers\cite{Chenarani.2022}.

Similar ranges for DIC are obtained using recently developed plant-based activated carbon materials\cite{SandovalGonzalez.2022,Souza.2021,CorreaNavarro.2020}. %
Another PFAS derivative, known as GenX, was recently investigated for its affinity for human serum albumin using the LIE method\cite{DelvaWiley.2021}, as done in this work. Four binding sites are reported, with similar affinities in the range of --$7.4$ to --$1.5$~\si{\kcal\per\mol}, despite its much larger size.

Commercially available activated carbon is commonly used at the latest stage in wastewater treatment and was investigated regarding its adsorption ability for diclofenac as well. The reported binding free energies are above\cite{Jodeh.2016,Petrovic.2020} $-2$~\si{\kcal\per\mol}, that is, weaker than the values obtained in this study.
Thus, the binding affinities of the ligands on the heterodimer obtained in this study are significant and comparable to those of other materials used for the filtration or purification of contaminated water.
\FloatBarrier
\section{Conclusions}

In this study, the only available hair keratin heterodimer in the literature, formed by the keratin protein pair K35/K85, was investigated in terms of its adsorption capability for the pharmaceutical diclofenac and the perfluorinated compound PFBS. Both types of substances are detectable in human hair samples, suggesting that human hair acts as an adsorbent. At least for diclofenac, melanin pigments are ruled out as an adsorbent which motivates that the keratin heterodimer is the active adsorbing component in human hair. An initial estimate 
of the heterodimer affinity was obtained using the VINA molecular docking program, suggesting multiple, high-affinity sites for both PFBS and DIC at the HEAD and TAIL regions, providing more positively charged and buried sites, compared to the open helical segments.

Therefore, further investigation was focused on the HEAD region by subjecting selected docking results to standard MD simulations of more than $1$~\si{\micro\second},  which revealed several long time stable sites for both ligands. The Gibbs free energy was measured by the LIE interaction method, and the molecular interaction fingerprint was determined to detect key interaction residues. In the HEAD region, four sites for diclofenac and two four PFBS were identified. In these cases, a strong interaction of the anionic ligands with at least one basic residue of the protein heterodimer was found.
However, all four stable sites of the ligand DIC included $\pi$--stacking or $\pi$--cation interactions that contribute to the high binding affinities.

The binding affinities obtained are higher for diclofenac compared to PFBS. This is attributed to the extended interaction possibilities of the diclofenac molecule. While other materials with higher adsorption affinities are studied in the literature, commonly used commercial activated carbon was shown to exhibit lower affinity for diclofenac. Therefore, we suggest further investigation of hair keratin as  a possible cheap and abundant adsorption material in water filtration applications.

Future work might address a higher order complex of keratin proteins, such as bundles of two or more dimers forming tetramers or microfibril structures in order to further investigate the accuracy of our predictions, while single dimer experimental validation could be inspired by the work of ref \citenum{Matsunaga.2013}.
Additional research may aim at investigating site-specific mutations to further improve the binding strength of the ligands, suggested to be conducted once experimentally validated structures of hair keratin heterodimers are available.
Also of interest is the study of binding capacity per dimer, in terms of the number of ligands bound, which requires a larger data base.

\FloatBarrier
\section*{Acknowledgments}
This work was supported by the Fraunhofer Internal Programmes under Grant No. MEF 835617. We gladly acknowledge supercomputer time on the Elwetritsch cluster at RHRK of RPTU Kaiserslautern, as well as on the  Beehive cluster of Fraunhofer ITWM.

\section*{AUTHORS}

\begin{itemize}
	
\item Ren\'e Hafner:  \href{https://www.orcid.org/0000-0003-3683-662X}{\textcolor{orcidlogocol}{\includegraphics[scale=1]{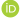}} orcid.org/0000-0003-3683-662X}

\item Nils Wolfgramm:  \href{https://www.orcid.org/0009-0004-9865-7250}{\textcolor{orcidlogocol}{\includegraphics[scale=1]{orcidlogo}} orcid.org/0009-0004-9865-7250}

\item Peter Klein: \href{https://orcid.org/0000-0002-5468-8889}{\textcolor{orcidlogocol}{\includegraphics[scale=1]{orcidlogo}}orcid.org/0000-0002-5468-8889}
 
\item Herbert Urbassek: \href{https://orcid.org/0000-0002-7739-4453}{\textcolor{orcidlogocol}{\includegraphics[scale=1]{orcidlogo}}orcid.org/0000-0002-7739-4453}

\end{itemize}

\section*{Data availability}
{The relaxed keratin structure along with the force field parameter files %
are made available at the Fordatis web server of the Fraunhofer Society.\cite{K35-K85-molmod} }
The data that support the findings of this study are available from the corresponding author upon reasonable request.

\bibliography{biosorb_molecular_citavi,biosorb_molecular_missing,biosorb-misc}
\clearpage

\end{document}